\documentclass[12pt,fleqn]{article}
\usepackage{amsmath} 
\usepackage{amssymb}

\makeatletter

\def\paragraph{\@startsection{paragraph}{4}{\z@}{+2.00ex plus
 +1ex minus +.2ex}{1.5ex plus .2ex}{\it\normalsize}}

\def\section{\@startsection {section}{1}{\z@}{+3.0ex plus +1ex minus
  +.2ex}{2.3ex plus .2ex}{\normalsize\bf\boldmath}}
\def\subsection{\@startsection{subsection}{2}{\z@}{+2.5ex plus +1ex
minus +.2ex}{1.5ex plus .2ex}{\normalsize\bf\boldmath}}
\def\subsubsection{\@startsection{subsubsection}{3}{\z@}{+3.25ex plus
 +1ex minus +.2ex}{1.5ex plus .2ex}{\normalsize\it}}

\expandafter\ifx\csname mathrm\endcsname\relax\def\mathrm#1{{\rm #1}}\fi

\newcounter{saveeqn}

\@addtoreset{equation}{section}

\newcount\@tempcntc
\def\@citex[#1]#2{\if@filesw\immediate\write\@auxout{\string\citation{#2}}\fi
  \@tempcnta\z@\@tempcntb\m@ne\def\@citea{}\@cite{\@for\@citeb:=#2\do
    {\@ifundefined
       {b@\@citeb}{\@citeo\@tempcntb\m@ne\@citea
        \def\@citea{,\penalty\@m\ }{\bf ?}\@warning
       {Citation `\@citeb' on page \thepage \space undefined}}%
    {\setbox\z@\hbox{\global\@tempcntc0\csname
b@\@citeb\endcsname\relax}%
     \ifnum\@tempcntc=\z@ \@citeo\@tempcntb\m@ne
       \@citea\def\@citea{,\penalty\@m}
       \hbox{\csname b@\@citeb\endcsname}%
     \else
      \advance\@tempcntb\@ne
      \ifnum\@tempcntb=\@tempcntc
      \else\advance\@tempcntb\m@ne\@citeo
      \@tempcnta\@tempcntc\@tempcntb\@tempcntc\fi\fi}}\@citeo}{#1}}

\def\@citeo{\ifnum\@tempcnta>\@tempcntb\else\@citea
  \def\@citea{,\penalty\@m}%
  \ifnum\@tempcnta=\@tempcntb\the\@tempcnta\else
   {\advance\@tempcnta\@ne\ifnum\@tempcnta=\@tempcntb \else
\def\@citea{--}\fi
    \advance\@tempcnta\m@ne\the\@tempcnta\@citea\the\@tempcntb}\fi\fi}

\def\asymp#1%
{\mathrel{\raisebox{-.4em}{$\widetilde{\scriptstyle #1}$}}}

\def\Nequal#1%
{\mathrel{\raisebox{-.5em}{$\stackrel{=}{\scriptstyle\rm#1}$}}}
\newcommand{\dsl}[1]{\not \hspace{-0.7mm}#1}
\def\dsl{\mathpalette\make@slash}
\def\make@slash#1#2{\setbox\z@\hbox{$#1#2$}%
  \hbox to 0pt{\hss$#1/$\hss\kern-\wd0}\box0}

\def\draftdate{\relax}
\def\mda{\relax}
\def\mua{\relax}
\def\mla{\relax}
\def\draft{
\def\thtystars{******************************}
\def\sixtystars{\thtystars\thtystars}
\typeout{}
\typeout{\sixtystars**}
\typeout{* Draft mode!
         For final version remove \protect\draft\space in source file *}
\typeout{\sixtystars**}
\typeout{}
\def\draftdate{\today}
\def\mua{\marginpar[\boldmath\hfil$\uparrow$]%
                   {\boldmath$\uparrow$\hfil}%
                    \typeout{marginpar: $\uparrow$}\ignorespaces}
\def\mda{\marginpar[\boldmath\hfil$\downarrow$]%
                   {\boldmath$\downarrow$\hfil}%
                    \typeout{marginpar: $\downarrow$}\ignorespaces}
\def\mla{\marginpar[\boldmath\hfil$\rightarrow$]%
                   {\boldmath$\leftarrow $\hfil}%
                    \typeout{marginpar: $\leftrightarrow$}\ignorespaces}
\def\Mua{\marginpar[\boldmath\hfil$\Uparrow$]%
                   {\boldmath$\Uparrow$\hfil}%
                    \typeout{marginpar: $\uparrow$}\ignorespaces}
\def\Mda{\marginpar[\boldmath\hfil$\Downarrow$]%
                   {\boldmath$\Downarrow$\hfil}%
                    \typeout{marginpar: $\downarrow$}\ignorespaces}
\def\Mla{\marginpar[\boldmath\hfil$\Rightarrow$]%
                   {\boldmath$\Leftarrow $\hfil}%
                    \typeout{marginpar: $\leftrightarrow$}\ignorespaces}
\overfullrule 5pt
\oddsidemargin -15mm
\marginparwidth 29mm
}

\def\stars{\strut\leaders\hbox{*}\hfill\strut}
\def\starline{\hfil\strut\hfil\hbox to \textwidth {\stars}\hfil}

\newcommand{\rwo}[2]{{\bar {#1}}^{\dot {#2}}}
\newcommand{\rwu}[2]{{\bar {#1}}_{\dot {#2}}}

\def\L{{\mathcal L}}

\def\F{{\mathcal F}}
\def\S{{\mathcal S}}
\def\O{{\mathcal O}}
\def\XS{{\scriptstyle \mathcal S}}

\def\G{\Gamma}
\def\Gcl{\Gamma_{\rm cl}}
\def\Gct{\Gamma_{\rm ct}}
\def\Gctsymrest{\Gamma_{\rm ct,restore}}
\def\Gctsym{\Gamma_{\rm ct,sym}}
\def\i{{\rm i}}

\def\a{\alpha}
\def\b{\beta}

\def\g{\gamma}
\def\d{\delta}
\def\e{\epsilon}
\def\l{\lambda}
\def\n{\nu}
\def\m{\mu}
\def\r{\rho}
\def\s{\sigma}

\def\w{\omega}

\newcommand{\SLASH}[2]{\makebox[#2ex][l]{$#1$}/}

\newcommand{\pslash}{\SLASH{p}{.2}}

\allowdisplaybreaks[3]

\begin{document}

\thispagestyle{empty}
\def\thefootnote{\fnsymbol{footnote}}
\setcounter{footnote}{1}
\null
\draftdate\hfill KA-TP-08-2003\\
\strut\hfill DESY-133-03\\
\strut\hfill MPP-2003-95\\
\strut\hfill hep-ph/0310191
\vfill
\begin{center}
{\Large \bf\boldmath
Restoration of supersymmetric Slavnov-Taylor and Ward identities in 
presence of soft and spontaneous symmetry breaking    
\par} \vskip 2.5em
\vspace{1cm}

{\large
{\sc I.\ Fischer$^1$\footnote{\it Now at the Institut f\"ur Theorie der Kondensierten Materie, Universit\"at Karlsruhe, D-76128 Karlsruhe, Germany}, W.\ Hollik$^2$, M.\ Roth$^2$ and D. St{\"o}ckinger$^3$} } \\[1cm]
$^1$ {\it Institut f\"ur Theoretische Physik, Universit\"at Karlsruhe \\
D-76128 Karlsruhe, Germany} \\[0.5cm]
$^2$ {\it Max-Planck-Institut f\"ur Physik 
(Werner-Heisenberg-Institut) \\
D-80805 M\"unchen, Germany}
\\[0.5cm]
$^3$ {\it Deutsches Elektronen-Synchrotron DESY \\
D-22603 Hamburg, Germany}
\par \vskip 1em
\end{center}\par
\vskip 2cm {\bf Abstract:} \par Supersymmetric Slavnov-Taylor and Ward 
identities are investigated in presence of soft and spontaneous symmetry
breaking. We consider an abelian model where soft supersymmetry breaking
yields a mass splitting between electron and selectron and triggers 
spontaneous symmetry breaking, and we derive corresponding identities 
that relate the electron and selectron masses with the Yukawa coupling. 
We demonstrate that the identities are valid in dimensional reduction 
and invalid in dimensional regularization and compute the necessary
symmetry-restoring counterterms.
\par
\vskip 1cm
\noindent
October 2003
\null
\setcounter{page}{0}
\clearpage
\def\thefootnote{\arabic{footnote}}
\setcounter{footnote}{0}

\section {Introduction}

In the past years, great progress has been made in the calculation of 
quantum corrections to precision observables. The calculation of 
such corrections necessitates the use of a regularization method in the 
intermediate steps of the renormalization procedure. 
Regularization schemes that preserve all or at least most symmetries of 
the underlying theory are most convenient. 
However, as far as supersymmetric gauge theories are concerned, a
regularization method that respects all symmetries and is mathematically 
consistent is not yet known: Dimensional regularization (DREG) 
\cite{DREG} breaks 
supersymmetry, while dimensional reduction (DRED) \cite{Siegel:1979wq}, a 
variation of DREG widely used in practical calculations, is inconsistent 
\cite{Inconsist} and thus cannot work at all orders. 

For quantization the basic symmetries of the theory are 
formulated as relations between renormalized Green functions. In order 
to test whether the symmetries are respected by the renormalization 
scheme or not, the corresponding symmetry identities have to be 
evaluated by an explicit calculation. 
If symmetry violations occur, the symmetry breakings have to be absorbed 
by so-called 
symmetry-restoring counterterms, a procedure often called algebraic 
renormalization (see Ref.\ \cite{ARbook} for an introduction).  

Supersymmetric gauge theories exhibit a few peculiarities that have to be 
addressed in the following. 

In practice, calculations are almost exclusively carried out in the
Wess-Zumino gauge. This yields a minimal number of unphysical particles
but makes it necessary to include the supersymmetry algebra in the BRS 
transformations. As a result, the supersymmetry transformations become
non linear, and the invariance of the action under supersymmetry 
transformations can no longer be expressed by Ward identities 
\cite{Breitenlohner:kh} but by more complicated Slavnov-Taylor 
identities \cite{White:ai,Maggiore:1995gr} (see also \cite{Ohl:2002jp}
for a recent discussion). 

An additional complication arises from the fact that supersymmetry is softly
broken in all phenomenologically relevant supersymmetric models to account 
for the mass splittings within supermultiplets. Therefore, it is necessary 
to include the description of soft supersymmetry breaking in the defining 
symmetry identities of the theory. This has been worked out in Refs.\ 
\cite{Girardello:1981wz,Maggiore:1996gg,Hollik:2000pa,Hollik:2002mv,Kraus:2002ri}. 
Furthermore, soft supersymmetry breaking triggers the spontaneous breaking of
internal symmetries. Both soft and spontaneous symmetry breaking 
generate mass terms in the Lagrangian.

Supersymmetric Slavnov-Taylor identities have been formulated for many 
physically relevant models: For generic supersymmetric Yang-Mills
theories in Refs.\ \cite{Maggiore:1995gr,Maggiore:1996gg,Hollik:2000pa}, for 
supersymmetric QED (SQED) in Ref.\ \cite{Hollik:1999xh}, and the Minimal
Supersymmetric Standard Model (MSSM) in Ref.\ \cite{Hollik:2002mv}.

However, the possible violations of these identities in DREG or DRED and
the necessary symmetry-restoring counterterms have only been
derived in very special cases: In Ref.\ \cite{Capper:ns} several
supersymmetric Ward identities for self energies have been shown to be
fulfilled in DRED, supersymmetric Slavnov-Taylor identities have been
investigated for SQED in Ref.\ \cite{Hollik:1999xh} and for SQCD with soft 
breaking in Ref.\ \cite{Hollik:2001cz}. In Ref.\ \cite{Hollik:2001cz} the 
symmetry-restoring counterterms for all gauge and gaugino interactions 
have been derived; thereby, it turned out that the presence of soft 
supersymmetry breaking terms have no influence on the determination of 
the counterterms.
Up to now, no explicit calculation of a supersymmetric Slavnov-Taylor
identities has been performed for cases where soft supersymmetry breaking is
directly relevant, e.g.\ sparticle masses.

The aim of the present work is to present such a calculation 
and to investigate the effect of soft 
supersymmetry breaking on the possible violations of the Slavnov-Taylor 
identity in DREG and DRED. 
Because of the complexity of the MSSM we choose a simplified 
model where the important features of the MSSM, i.e.\ soft supersymmetry
breaking and spontaneous breaking of an internal symmetry, are retained. 

The outline of this article is as follows: In Section 2 we define the
model and present its symmetries, in particular the
Slavnov-Taylor identity. Section 3 contains information of general
importance for the explicit calculation of symmetry identities. We
discuss how symmetry identities are violated in DREG and DRED
and study when soft supersymmetry breaking becomes relevant in 
Slavnov-Taylor identities. In Section 4 we consider two symmetry
identities for vertex functions as an example. We present the results 
of an analytical evaluation at one-loop level for arbitrary 
(on- or off-shell) momenta of the external particles, derive the violation 
of the identities and the corresponding symmetry-restoring counterterms. 
Our conclusions are presented in Section 6.

\section {The model and its symmetries}
\label{model}

\subsection{Particle content and Lagrangian}

The MSSM is a supersymmetric gauge theory with soft supersymmetry
breaking and spontaneous breaking
of an internal symmetry. Our aim is to construct a model that retains
these essential features of the MSSM concerning symmetry breaking and 
is maximally simplified. Hence we restrict ourselves to 
SQED extended by an Higgs sector, i.e.\ we reduce the gauge group and the
matter content to the ones of SQED (U(1) gauge group; two
chiral supermultiplets $\Phi_{1,2}$ with charges $Q_{1,2}=\mp1$
corresponding to the left- and right-handed electron) extended
by an additional uncharged chiral multiplet $\Phi_3$. The field $\Phi_3$
takes the role of a Higgs field and will acquire a vacuum expectation value
(VEV). In addition to U(1) gauge invariance, we require invariance
under a continuous $R$-transformation with $R$-charges $n=2/3$ for the
chiral multiplets $\Phi_{1,2,3}$.

The corresponding supersymmetric Lagrangian reads in superfield notation
\begin{eqnarray}
\L_{\rm susy} & = & \int {\rm d}^4\theta 
\sum_{i=1,2,3}\Phi_i^\dagger e^{eQ_iV}\Phi_i
\nonumber\\
&&{}
+\left[\int {\rm d}^2\theta\left(\frac14 W^\a W_\a + W(\Phi)\right) +{\rm h.c.}\right],
\end{eqnarray}
where $V$ denotes the gauge superfield and $W_\a$ is the corresponding
field strength superfield. The superpotential $W(\Phi)$ takes
the form
\begin{eqnarray}
W(\Phi) & = & \frac16\sum_{ijk}y_{ijk} \Phi_i\Phi_j\Phi_k
\end{eqnarray}
with the totally symmetric parameters $y_{ijk}$. Because of gauge
invariance there are only two independent parameters 
$y_{123}=y_{213}=y_{132}=\ldots$ and $y_{333}$; the others are zero.
Owing to the specific choice of the $R$-charges no dimensionful
parameters are possible in $\L_{\rm susy}$.

The components of the gauge superfield $V$ are the photon and 
photino fields $(A^{\m}$, $\l_{\a}$, $\bar{\l}^{\dot{\a}})$, 
while the chiral multiplets
$\Phi_{1,2,3}$ contain the component fields $(\phi_L, \psi_L^{\a} )$, 
$(\phi_R, \psi_R^{\a} )$, $(H, \tilde H^{\a} )$, respectively.
The Weyl spinors are combined into 4-component spinors for the photino,
electron, and Higgsino as follows,
\begin{equation} 
\tilde\g = \left( \begin{array}{c} -\i
      \l_{\a} \\ \i \rwo{\l}{\a} \end{array} \right),\quad
\Psi = \left(
      \begin {array}{c}{\psi_L}_{\a} \\ {\bar\psi}_R^{\dot\a} \end
 {array} \right),\quad
\hat H = \left( \begin {array}{c}{\tilde H_{\a}} \\ \rwo{\tilde H}{\a}
  \end {array} \right).
\end{equation}

Soft breaking of supersymmetry is introduced via power-counting
renormalizable and supersymmetric couplings to a chiral (``spurion'')
superfield $\eta$ of dimension 0
\cite{Hollik:2000pa,Hollik:2002mv}:
\begin{eqnarray}
\L_{\rm soft} & = & \int {\rm d}^4\theta
\sum_{i=1,2,3}\tilde{M}^2_{ii}\eta^\dagger\eta\Phi_i^\dagger e^{2eQ_iV}\Phi_i
\nonumber\\
&&{}
+\left[\int {\rm d}^2\theta \left(
     \frac 12 \tilde{M}_\lambda\eta W^\a W_\a
   + \sum_{ijk}\tilde{A}_{ijk}\eta\Phi_i\Phi_j\Phi_k\right) + {\rm h.c.}\right]
\label{Lsoft}
\end{eqnarray}
with the soft supersymmetry breaking parameters $\tilde{M}_{ii}$, 
$\tilde{M}_\lambda$, and $\tilde{A}_{ijk}$.
The component fields of $\eta$ are called $(a,\chi^\a,f+f_0)$ where the
auxiliary $f$-component has a constant shift $f_0$. 
We define the 4-spinor 
\begin{equation}
\label{Chi}
\Xi=\left(\begin{array}{c}{\chi_\a}\\{\rwo{\chi}{\a}}\end{array}\right)
\end{equation}
for later use. This $\eta$-multiplet 
is treated as an external field that does not propagate. As long
as $\eta$ transforms as a chiral multiplet, $\L_{\rm soft}$ is
supersymmetric. If $\eta$ is replaced by the constant shift,
\begin{equation}
\eta \to \theta\theta f_0,\quad
(a,\chi,f+f_0) \to (0,0,f_0),
\end{equation}
$\L_{\rm soft}$ breaks supersymmetry softly.
At the same time, continuous $R$-invariance is explicitly
broken.\footnote{The remaining discrete
  symmetry is simply a parity transformation where all spinors have
  parity $-1$ and all other fields $+1$. Invariance under this
  transformation is a trivial consequence of Lorentz invariance and
  does not correspond to $R$-parity. Anyway, the model is
  invariant under $R$-parity where the Standard-Model like fields (photon,
  electron, Higgs boson) have parity $+1$ and their superpartners $-1$.}

Interactions like $\eta W^\alpha W_\alpha$ or $\eta \Phi^3$ are not 
present in power-counting renormalizable supersymmetric models that 
contain only $\mathrm{dimension}=1$ chiral superfields. Such
interactions have not been considered so far in checks of symmetry 
identities at the regularized level.
The presence of such spurion interactions makes the study of symmetry
identities particularly interesting (see Section \ref{MassRelations})
since they might lead to additional violations in DREG or even in DRED.

The dimensionful parameters in $\L_{\rm soft}$ lead to a scalar
potential with a minimum at a finite VEV, $\langle H\rangle =v$. 
As a consequence, $R$-invariance is spontaneously broken and the scalar and
fermionic particles become massive. In the following we split off the 
VEV $v$ from the Higgs field $H$ by the replacement
\begin{equation}
H\to H+v.
\end{equation}

Soft and spontaneous symmetry breaking lead, as in the MSSM, to a relation 
between the Yukawa coupling $y_{123}$ for the $H\psi_L\psi_R$- or 
$\tilde{H}\psi_L\phi_R$ interactions, the electron mass 
\begin{equation}
m_e = v y_{123},
\end{equation}
and the selectron mass matrix
\begin{equation}
\label{selectronmass}
M_{\phi}^2 = \left(\begin{array}{cc}
             m_e^2+M_{11}^2  &  y_{333}v m_e/2+6v A_{123} \\
             y_{333}v m_e/2+6v \tilde A_{123}  &
             m_e^2+M_{11}^2\end{array}\right)
\end{equation}
with $M_{11}^2=f_0^2\tilde{M}^2_{11}$ and $A_{123}=f_0\tilde{A}_{123}$.

In contrast to the MSSM, there are no gauge bosons in this model that couple to
Higgs fields. Therefore, gauge invariance is not broken spontaneously,
and there are no massive gauge bosons. A second consequence is that 
no $D$-terms $\propto e^2 v^2$ contribute to the mass matrix
$M_{\phi}^2$ in our model.

\subsection{Symmetry identities}

In order to complete the definition of our model as a renormalized 
quantum field theory, we rewrite in the following the symmetry
requirements in form of Slavnov-Taylor and Ward identities.

Because of the ambiguity inherent in the regularization and
renormalization procedure, specifying the Lagrangian is
not sufficient to define the model at higher orders. Instead, the
model has to be defined by requiring that $\Gamma$, the renormalized
effective action or generating functional of one-particle irreducible
Green functions, has the same symmetry properties as the classical
action $\Gamma_{\rm cl}$. For this purpose, the symmetries have to be
formulated in terms of well-defined identities for $\Gamma$ that have to be
fulfilled in all orders. The structure of such identities for the case of 
supersymmetric gauge theories, quantized in the Wess-Zumino gauge, has 
been studied in Refs.\ 
\cite{White:ai,Maggiore:1995gr,Maggiore:1996gg,Hollik:2000pa,Hollik:2002mv,Hollik:1999xh}.

Gauge invariance and supersymmetry are treated simultaneously by
introducing BRS trans\-for\-mations comprising gauge symmetry,
supersymmetry, and translations. This requires three kinds of ghost fields: one
Faddeev-Popov ghost field corresponding to the gauge transformations $c(x)$ 
(a fermionic scalar), a supersymmetry ghost $\e_\a$ (a bosonic
spinor), and a translational ghost $\w^\m$ (a fermionic vector).  
As supersymmetry is global, neither the supersymmetry ghost nor the
translational ghost are dynamical fields. We refer to Appendix
\ref{brs} for the explicit form of the BRS transformations. 

Generally, if symmetry transformations of the classical action are 
non linear in propagating fields, not only $\Gamma$ receives loop
corrections but also the field transformations themselves. 
In our case, the non-linear BRS transformations $s \varphi_i$ (see Appendix
\ref{brs}) receive loop corrections and have to be renormalized. This
is usually done by coupling them to external sources 
$Y_{\varphi_i}$ and including them in the effective action.  

The gauge-fixing and ghost terms can be conveniently written as a
total BRS variation. For this purpose, the antighost $\bar c$ and an
auxiliary field $B$ are introduced with suitable BRS variations.
The complete form of the effective action in lowest order,
the classical action $\G_{\rm cl}$, is given in Appendix \ref{lagrangian}. 

The classical action satisfies the Slavnov-Taylor identity
$\S(\G_{\rm cl})=0$ with the Slavnov-Taylor operator (given for Weyl spinors):
\begin{eqnarray} 
\label{DerSTO}
\S & = & \S_0 + \S_{\rm soft}, \\
\S_0(\F) & = & \int {\rm d}^4x\, \left(sA^{\m} \frac{\d\F}{\d A^{\m}} + sc
  \frac{\d\F}{\d c} + s\bar c \frac{\d\F}{\d \bar c} + sB \frac{\d\F}{\d
  B} \right. \nonumber\\
& & \quad + \frac{\d\F}{\d Y_{\l\a}}\frac{\d\F}{\d\l^{\a}} +
  \frac{\d\F}{\d Y_{\bar\l}^{\dot\a}}\frac{\d\F}{\d\bar\l_{\dot\a}} \nonumber\\
& & \quad + \frac{\d\F}{\d Y_{\phi_{L}}}\frac{\d\F}{\d \phi_{L}} + 
  \frac{\d\F}{\d Y_{\phi_{L}^{\dagger}}}\frac{\d\F}{\d \phi_{L}^{\dagger}} + 
  \frac{\d\F}{\d Y_{\psi_{L\a}}}\frac{\d\F}{\d \psi_{L}^{\a}} + 
  \frac{\d\F}{\d Y_{\bar\psi_{L}}^{\dot\a}}\frac{\d\F}{\d\bar\psi_{L\dot\a}} + 
  (_{L \rightarrow R}) \nonumber\\
& & \quad \left. + s(H + v)\frac{\d\F}{\d H} + s(H^{\dagger} + v) \frac{\d\F}{\d
  H^{\dagger}} + \frac{\d\F}{\d Y_{\tilde H}}\frac{\d\F}{\d \tilde H} + 
  \frac{\d\F}{\d Y_{\bar{\tilde H}}}\frac{\d\F}{\d \bar{\tilde H}} \right) \nonumber\\
& & + s\e^{\a} \frac{\partial\F}{\partial\e^{\a}} +
  s\bar\e_{\dot\a} \frac{\partial\F}{\partial\bar\e_{\dot\a}} +
  s\w^{\n} \frac{\partial\F}{\partial\w^{\n}}, \\
\S_{\rm soft}(\F) & = & \int {\rm d}^4x\, \bigg( \bigg. sa
\frac{\d\F}{\d a} + sa^{\dagger}
 \frac{\d \F}{\d a^{\dagger}} + s \chi^{\a}
 \frac{\d \F}{\d \chi^{\a}} + s \rwu{\chi}{\a}
 \frac{\d \F}{\d \rwu{\chi}{\a}} \nonumber\\
& & \quad + s(f + f_0) \frac{\d \F}{\d f} + s(f^{\dagger} + f_0)
 \frac{\d \F}{\d f^{\dagger}} \bigg. \bigg).
\end {eqnarray}
We abbreviate this by
\begin {equation}
\S(\F) \equiv \int {\rm d}^4 x \, \left(\sum \limits_{i} s \varphi^\prime_i \frac{\d
  \F}{\d \varphi^\prime_i} + \frac{\d \F}{\d Y_{\varphi_i}} \frac{\d \F}{\d
  \varphi_i}\right),
\end {equation}
where the fields with linear BRS transformations are denoted by
$\varphi_i^{\prime}$ and the fields with non-linear BRS transformations are
denoted by $\varphi_i$.
The corresponding linearized Slavnov-Taylor operator reads
\begin {equation}
{\scriptstyle \mathcal S}_{\F} = \int {\rm d}^4 x \, \left(\sum \limits_{i}
s \varphi^\prime_i \frac{\d}{\d \varphi^\prime_i} + 
\frac{\d \F}{\d Y_{\varphi_i}} \frac{\d}{\d \varphi_i} +
\frac{\d \F}{\d \varphi_i} \frac{\d}{\d Y_{\varphi_i}} \right).
\end {equation}
It satisfies
\begin{equation}
\S(\F+\delta\F)=\S(\F)+{\scriptstyle \mathcal S}_{\F} \delta\F+ {\cal
  O}(\delta\F^2).
\end{equation}

In addition to supersymmetry and gauge invariance, we require
invariance under continuous $R$-transformations with the $R$-charges
$2/3$ for $\Phi_{1,2,3}$ and 0 for $\eta$. The corresponding Ward
operator is given by 
\begin{eqnarray}
\label{Rsym}
{\mathcal W} (\F) & = & 
\int {\rm d}^4 x \, \i\left(\sum_i  n_{\varphi'_i}(\varphi'_i+v_i)
\frac{\d\F}{\d\varphi'_i}
+n_{\varphi_i}\varphi_i\frac{\d\F}{\d\varphi_i}
+n_{Y_{\varphi_i}}Y_{\varphi_i}\frac{\d\F}{\d Y_{\varphi_i}}\right).
\end{eqnarray}
Here $v_i$ denotes the finite VEV of the field $\varphi'_i$
and is only non-zero for $\varphi'_i=H,H^\dagger,f,f^\dagger$.
The $R$-charges are listed in Table \ref{tab:R}.
$R$-symmetry is spontaneously and explicitly broken as can be seen by the
appearance of $v$ and $f_0$ in Eq.\ (\ref{Rsym}).
\begin{table}[t]
\label{tab:R}
\begin{center}
\begin{tabular}{|c|cccccccccccc|}
\hline 
Fields & $A_\m$ & $\l^{\a}$ & $\phi_L$ & $\phi_R$ & $\psi_L^{\a}$ &
$\psi_R^{\a}$ & $H$ & $\tilde H^{\a}$ & $\e^{\a}$ & $a$ & $\chi^{\a}$ & $\hat f$\\[1mm]
\hline
$R$-charges & 0 & $1$ & $2/3$ & $2/3$ & $-1/3$ & $-1/3$ & $2/3$ &
$-1/3$ & $1$ & 0 & $-1$ & $-2$ \\[1mm]
\hline
\end {tabular}
\end {center}
\caption{$R$-charges of the component fields}
\end{table}

To summarize, the model is defined by its field content and the following
conditions and symmetry requirements on the effective action $\Gamma$:
\begin{itemize}
\item Slavnov-Taylor identity and nilpotency of $\XS_\G$:
\begin{eqnarray}
\label{DIEsti}
{\mathcal S}(\G) & = & 0, \\
\label{nil}
{\scriptstyle \mathcal S}_{\G}^2 A^{\m} & = & 0.
\end{eqnarray}
This identity contains gauge invariance, supersymmetry, and
translational invariance. The abelian gauge group makes it necessary to 
impose an additional condition (\ref{nil}) to guarantee the nilpotency
of the Slavnov-Taylor operator \cite{Hollik:2002mv,Hollik:1999xh}. 
\item 
Gauge-fixing condition, ghost and anti-ghost equation, and 
translational ghost equation:
\begin{equation} 
\label{rest}
\frac{\d\G}{\d B} = \frac{\d\G_{\rm cl}}{\d B}, \quad
\frac{\d\G}{\d \bar c} = \frac{\d\G_{\rm cl}}{\d \bar c}, \quad
\frac{\d\G}{\d c} = \frac{\d\G_{\rm cl}}{\d c}, \quad
\frac{\d\G}{\d \w^{\m}} = \frac{\d\G_{\rm cl}}{\d \w^{\m}}. 
\end{equation}
These equations hold since they are linear in propagating fields, and
they express the non-renormalization of the gauge-fixing term 
and the usual QED-Ward identity.
\item Ward identity for $R$-symmetry:
\begin{equation} 
\label{DIEwi}
{\mathcal W}(\G) = 0.
\end{equation}
\item 
Global symmetries: We require that $\G$ is Lorentz invariant, 
bosonic, and electrically neutral. Furthermore, $\G$ has to be invariant 
under the discrete symmetries $C$, $CP$, and must not possess a ghost charge.
\item The physical content of $\G$ is given in the limit 
\begin{equation}
\label{physlimit}
\G_{\rm phys} = \G |_{a = \chi = f = 0}.
\end{equation}
In this limit, supersymmetry is softly broken.
\end{itemize}

\section{Renormalization and symmetry-restoring counterterms}
\label{sec:Renormalization}

At higher orders, the symmetry identities are generally violated at
the regularized level. They have to be restored by adding
suitable counterterms, so-called symmetry-restoring counterterms. 
This section contains
useful information on how the symmetries can be broken 
in dimensional schemes and how the symmetry-restoring counterterms can be
calculated. The symmetry identities determining these counterterms 
are classified using power-counting arguments and $R$-invariance.

\subsection{Structure of counterterms and symmetry-breaking terms}
\label{secStructure}
The calculation of $\Gamma$ at higher orders is an inductive
process. If the classical action $\G_{\rm cl}$ and the counterterms up
to order $\hbar^{n-1}$, $\G_{\rm ct}^{(\le n-1)}$, are known, the
regularized effective action $\G_{\rm reg}^{(\le n)}$ can be
calculated up to order $\hbar^n$. The renormalized effective action up
to this order is obtained by adding the counterterms 
$\G_{\rm ct}^{(n)}$:
\begin{equation}
\G^{(\le n)}=\G^{(\le n)}_{\rm reg}+\G_{\rm ct}^{(n)}.
\end{equation}
The counterterms are necessary to cancel ultra-violet divergences and to
restore the symmetry identities. They can be split into a symmetry-restoring
part and a part that does not interfere with the symmetries,
\begin{equation} \label{Gct}
\G_{\rm ct}^{(n)}=\Gctsymrest^{(n)}+\Gctsym^{(n)}.
\end{equation}
$\Gctsym$ contains the usual counterterms corresponding to
multiplicative renormalization of the parameters and fields.

We now focus on 
symmetry violations of Slavnov-Taylor identities induced 
by the regularization scheme. Assuming that the Slavnov-Taylor identity
is valid at the order $\hbar^{n-1}$, $\S(\G^{(\le   n-1)})=0$, we have
\begin{equation}
\S(\G_{\rm reg}^{(\le n)})=\Delta^{(n)}.
\label{SGreg}
\end{equation}
If the symmetry is free of anomalies, like in our model, $\Delta^{(n)}$ can
be absorbed by symmetry-restoring counterterms satisfying
\begin{eqnarray}
\XS_{\Gcl}\Gctsymrest^{(n)}& = & - \Delta^{(n)},\\
\S(\G_{\rm reg}^{(\le n)}+\Gctsymrest^{(n)}) & = & 
0 + {\cal O}(\hbar^{n+1}).
\end{eqnarray}

The symmetric counterterms $\Gctsym^{(n)}$ satisfy by definition
\begin{equation}
\XS_{\Gcl}\Gctsym^{(n)} = 0.
\end{equation}
They cancel divergences in $\G_{\rm reg}^{(\le n)}+\Gctsymrest^{(n)}$,
and their finite parts correspond to the free parameters of the model
and are fixed by suitable renormalization
conditions. Without loss of generality, we can require 
that $\G^{(n)}_{\rm ct, restore}$ does not contribute to
those vertex functions of order $\hbar^n$ on which renormalization 
conditions are imposed. The separation (\ref{Gct}) is then unambiguous.

In the following, we list several important properties of the
symmetry-breaking terms $\Delta^{(n)}$.
\begin{enumerate}
\item 
The Quantum Action Principle \cite{QAP} requires $\Delta^{(n)}$
to be a local, power-counting renormalizable polynomial in the fields
of ghost number +1. Furthermore, using the algebraic nilpotency
properties of the operators $\XS_\G$ and $\S(\G)$, 
one can derive the condition 
\begin{equation}
\XS_{\Gcl}\Delta^{(n)}=0.
\end{equation}
These properties are important in algebraic proofs of the absence of
anomalies and the renormalizability \cite{White:ai,Maggiore:1995gr,BRS}.
\item \label{itemRAP}
In DREG as defined in Refs.\ \cite{HV,Breitenlohner:hr} (HVBM scheme) 
the Quantum Action Principle can be made more precise (``regularized 
action principle''). As shown in Ref.\ \cite{Breitenlohner:hr}, the 
breaking $\Delta$ of any given symmetry identity can be directly computed:
\begin{equation} 
\label{bm1}
\Delta =
\left[\int {\rm d}^D x \, \d \L_{\rm cl+ct} \right] \cdot \Gamma, 
\end{equation}
where $[{\cal O}]\cdot\Gamma$ denotes an insertion of the operator $\cal O$ 
and $\int {\rm d}^D x \, \d\L_{\rm cl+ct}$ is the variation of the
classical action and counterterms under the
corresponding symmetry transformations in $D$ dimensions. E.g.\ for the
cases of the Slavnov-Taylor identity and $R$ Ward identity it is given by
$\S(\Gamma_{\rm cl+ct})$ and ${\mathcal W}\Gamma_{\rm cl+ct}$
evaluated in $D$ dimensions. For a symmetry identity that is valid at 
the classical level, $\d \L_{\rm cl}$ vanishes in four dimensions and 
has the form
\begin{equation} 
\label{bm2}
\delta {\mathcal L}_{\rm cl} = \O (D-4, \hat g^{\m\n}),
\end{equation}
where $\hat g^{\m\n}$ is the $(D-4)$-dimensional part of the
metric tensor. There are three important consequences for the one-loop
breaking $\Delta^{(1)}$:
\begin{itemize}
\item At the one-loop level, Feynman diagrams contain only one insertion of
$\d {\mathcal L}_{\rm cl}$. Since the divergences of the diagrams are
at most of the order of $1/(D-4)$, the breaking term $\Delta^{(1)}$ is
finite in the limit $\hat{g}^{\mu\nu}\to0$, $D\to4$. 
\item The variation $\delta {\mathcal L}_{\rm cl}$ is a polynomial in 
coupling constants, masses, kinematic variables, and fields.
The divergences of one-loop integrals have the same structure. Hence,
the breaking terms $\Delta^{(1)}$ are not only polynomials in 
fields and momenta, but even in mass parameters.
Particularly $\Delta^{(1)}$ contains no logarithms of mass parameters. 
\item
Moreover, the Quantum Action Principle becomes rather simple at the one-loop
level since counterterms appear only in tree diagrams and do not 
contribute to loops. In higher orders, the interplay 
of lower-order counterterms and loop contributions is crucial to ensure 
that the breaking $\Delta^{(n)}$ is local.  
\end{itemize}
\item The statements of Item \ref{itemRAP} are even valid in the naive
version of DREG, where an anticommuting $\g_5$ and accordingly 
${\rm Tr}(\g_5 \g^\mu\g^\nu\g^\rho\g^\sigma)=0$ are used. 
Of course, naive DREG is problematic since the limit
$D\to4$ of finite quantities such as 
${\rm Tr}(\g_5 \g^\mu\g^\nu\g^\rho\g^\sigma)=0$ does not agree with the 
four-dimensional result, but nevertheless this scheme is useful in
practical calculations where such traces do not appear. 

Other practically useful schemes are DRED or DREG where an anticommuting 
$\g_5$ is used and ${\rm Tr}(\g_5\g^\mu\g^\nu\g^\rho\g^\sigma)$ is set to its
four-dimensional value at the same time. However, in these schemes the
regularized action principle is not necessarily valid because they are
mathematically inconsistent, i.e.\ one initial expression can lead to
several, disagreeing results depending on the order of the calculational steps.
The results obtained in such inconsistent schemes are in agreement
with all physical requirements and can be used if they differ only by
local contributions of dimension $\le4$ from the results obtained in a
consistent scheme. 

\item In the HVBM scheme, $R$-invariance is broken at the regularized
level because $\int{\rm d}^Dx\,\d\L_{\rm cl}={\mathcal W}\Gcl\ne0$ in $D$ 
dimensions owing to the non-anticommuting $\g_5$. In naive DREG, 
on the other hand, $R$-invariance is fulfilled on the 
regularized (one-loop) level since 
$\int {\rm d}^Dx\,\d\L_{\rm cl}={\mathcal W}\Gcl=0$ in $D$ 
dimensions. If only $R$-invariant counterterms are added, 
$\int{\rm d}^Dx\,\d\L_{\rm cl+ct}={\mathcal W}\G_{\rm cl+ct}=0$
and $R$-invariance is valid without introducing symmetry-restoring 
counterterms. In both versions of DREG, gauge invariance 
and Eqs.\ (\ref{rest}) are not violated by the regularization scheme.

Therefore, when naive DREG is used, the symmetry-restoring counterterms 
$\Gctsymrest$ have to satisfy by themselves all the Eqs.\ (\ref{rest}) 
and (\ref{DIEwi}), in
particular they are $R$-invariant. The breaking of the Slavnov-Taylor
identity $\S(\G)=\Delta$ is restricted accordingly, in particular
\begin{equation}
{\mathcal W}\Delta=0.
\end{equation}
\end{enumerate}

\subsection{Calculating the symmetry-restoring counterterms}
\label{sec:classification}

Our aim in this subsection 
is to sketch the calculation of the counterterms
$\Gctsymrest$ that restore the Slavnov-Taylor identity. Lorentz and 
$R$-invariance are assumed to be respected by the regularization scheme. 
Several strategies for such calculations have been
proposed in the literature: In Ref.\ \cite{Grassi}, Taylor expansions in
the momenta are used to generate universal, regularization-independent
counterterms in an intermediate step 
of the calculation of symmetry restoring counterterms. 
In Ref.\ \cite{Sanchez}, the breaking
$\Delta$ is computed directly using the regularized action principle
(\ref{bm1}) and the equation $\Delta=-\XS_{\Gcl}\Gctsymrest$ is solved
explicitly.

The strategy
we use has often been applied  in the literature 
(see e.g.\ Refs.\ \cite{Capper:ns,Hollik:1999xh,Hollik:2001cz}):
Slavnov-Taylor identities of the form 
\begin{equation}
\label{master}
0 \stackrel{!}{=}
\left. \frac{\d \S(\G)}{\d\varphi_{k_1}\ldots\d\varphi_{k_N}}\right|_{\varphi
  = 0}
\end{equation}
are derived and the appearing Green functions are evaluated in the 
chosen regularization scheme. 
Without adding counterterms, the identities 
are in general violated. To restore these identities,
symmetry-restoring counterterms are determined in such a way that 
all Slavnov-Taylor identities are fulfilled.

Generically, the identity (\ref{master}) is a sum of terms of the form
\begin{equation}
\G_{{\cal M} Y_{\varphi_i}}\G_{\tilde{\cal M}\varphi_i},\quad 
\frac{\d\int {\rm d}^4x \, s\varphi_i'}{\d {\cal M}}\G_{\tilde{\cal M}\varphi_i'}
\end{equation}
with monomials ${\cal M,\tilde{M}}$ satisfying 
${\cal M\tilde{M}}=\varphi_{k_1}\ldots\varphi_{k_N}$. When these products
are evaluated at the one-loop level, Eq.\ (\ref{master}) becomes a
linear expression
in one-loop Feynman integrals and also in 
one-loop 
counterterms. The role of the symmetry violations and the counterterms 
is emphasized by rewriting $\S(\G)=0$ as
\begin{eqnarray}
\XS_{\Gcl}\Gct^{(1)} = -\S(\G_{\rm reg}^{(1)}) = - \Delta^{(1)}.
\end{eqnarray}
Taking the derivative $\d/(\d\varphi_{k_1}\ldots\d\varphi_{k_N})$,
identity (\ref{master}) can be equivalently written as
\begin{equation}
\left. \frac{\d\XS_{\Gcl}\Gct^{(1)}}{\d\varphi_{k_1}\ldots\d\varphi_{k_N}}\right|_{\varphi
  = 0} 
= - \left. \frac{\d\Delta^{(1)}}{\d\varphi_{k_1}\ldots\d\varphi_{k_N}}\right|_{\varphi
  = 0}
.
\label{master2}
\end{equation}
The breaking term $\Delta^{(1)}$ can be expanded in a basis of
monomials,
\begin{equation}
\Delta^{(1)}=\sum_j a_j\Delta_j.
\end{equation}
$\Delta_j$ denotes a monomial, i.e.\ a product of fields and
derivatives with ghost number $+1$ and dimension $\le4$ which are 
Lorentz and $R$-invariant. Each monomial $\Delta_j$
corresponds to one identity of the form (\ref{master2}) with the
fields $\varphi_{k_1}\ldots\varphi_{k_N}$ taken from
$\Delta_j$. There are in general several monomials containing the same
fields $\varphi_{k_1}\ldots\varphi_{k_N}$ but differing e.g.\ in the number of
derivatives. One specific monomial can be extracted out of Eq.\ 
(\ref{master2}) by taking only those terms from Eq.\ (\ref{master2})
that have the same momentum dependence as $\Delta_j$ (after Fourier transform
in momentum space). 

If all identities corresponding to the monomials
$\Delta_j$ are satisfied by a suitable choice of counterterms
$\Gct^{(1)}$, the Slavnov-Taylor identity is restored. In this way,
the restrictions on the $\Delta_j$ imply restrictions on the
identities that have to be considered. In the following, we present a
classification of the $\Delta_j$ and discuss the corresponding
identities. 

The criteria of this classification are based on power counting
and $R$-invariance. The fields appearing in $\Delta_j$ are
denoted by $\varphi_{k_1}\ldots\varphi_{k_N}$ and the power-counting
dimension of $\varphi_{k_1}\ldots\varphi_{k_N}$ is called $d$. We set
the power-counting dimensions of the dynamical fields to their 
space-time dimensions. In general, we
know that each monomial is a Lorentz scalar of ghost number $+1$ that
does not depend on the ghosts $c$, $\omega^\mu$, and on the fields
$B$, $\bar c$ owing to manifest gauge invariance and the identities
(\ref{rest}). Hence, $\Delta_j$ must
contain at least one $\epsilon_\a$ or $\bar\epsilon^{\dot\a}$ ghost.
In Ref.\ \cite{Hollik:2000pa} it has been shown that the Slavnov-Taylor 
identity can be considered in the limit $a=\chi=0$ without losing information
on Green functions that do not depend on spurion and $Y$-fields. 
Along the same lines, it can be
shown that in our case, where $R$-invariance is respected by the 
regularization scheme, even $f=0$ can be used \cite{Kraus:2002ri}. 
Hence only $\Delta_j$ without $a,\chi,f$ are needed.
\begin{description}
\item{${\rm dim}\Delta_j=4$}: Since $R$-symmetry relates
monomials with non-vanishing $R$-charge to higher-dimensional ones, but
$\Delta$ contains only monomials of dimension $\le4$, $\Delta_j$ must
have $R$-charge zero. The number of space-time derivatives
in $\Delta_j$ is $4-d$ and, thus, the relevant identity reads
\begin{equation}
0\stackrel{!}{=}\left[\left.
\frac{\d\S(\G)}{\d\varphi_{k_1}\ldots\d\varphi_{k_N}}
\right|_{\varphi = 0}\right]_{{\rm Terms}\propto p^{4-d}},
\label{dim4}
\end{equation}
where $p$ denotes generically all appearing momenta. 
From power counting it is easy to see that every term in this identity
can behave at most as $p^{4-d}$ (up to logarithms). Hence the relevant
terms are the leading terms in the momenta. In order to extract the
necessary information, the identity can be evaluated in the limit
$p\to\infty$ and all mass parameters can be neglected. This is an
enormous simplification. 

On the other hand, only
counterterms of $\mathrm{dimension}=4$ contribute and can be
determined using identities of this kind. Thus, no symmetry-breaking
effects do contribute to Eq.\ (\ref{dim4}). Hence, $f_0$ and $v$ can be
neglected. Such identities and these simplifications have been
discussed and used in Ref.\ \cite{Hollik:2001cz} to derive the counterterms
to all $\mathrm{dimension}=4$ gluon and gluino interactions.
\item ${\rm dim}\Delta_j=3$: $\Delta_j$ need not have $R$-charge
zero. It can originate from a term such as $(f+f_0)\Delta_j$ or
$(H+v)\Delta_j$ which has dimension 4 and is $R$-invariant. The
possibilities for the $R$-charge $n(\Delta_j)$ are: $0,\pm n_H,\pm
n_f$. They will be discussed later on. Since the relevant identity is
\begin{equation}
0\stackrel{!}{=}\left[\left.
\frac{\d\S(\G)}{\d\varphi_{k_1}\ldots\d\varphi_{k_N}}
\right|_{\varphi = 0}\right]_{{\rm Terms}\propto p^{3-d}},
\label{dim3}
\end{equation}
the vertex functions have to be evaluated up to the subleading
order in $p$ and the masses cannot be neglected. Nevertheless, one
simplification is possible: Since $\Delta^{(1)}$ is a polynomial in
the mass parameters in DREG (see Item
\ref{itemRAP} in Section \ref{secStructure}), it is
sufficient to evaluate the leading terms in the Taylor expansion in
dimensionful couplings and masses (e.g.\ using the mass-insertion
method), neglecting $\log m$-terms.
This simplification is particularly valuable in models like
the MSSM with many masses and mixing parameters.

Let us now go through the cases for $n(\Delta_j)$ and discuss the
effects of symmetry breaking. If $n(\Delta_j)=- n_f$ and if
$\Delta_j$ contains $\epsilon_\a$, i.e.\ $\varphi_k=\epsilon_\a$, soft
supersymmetry breaking appears in the Slavnov-Taylor identity. The
relevant term in $\d\S(\G)/(\d\varphi_{k_1}\ldots\d\varphi_{k_N})=0$ reads
\begin{equation}
\frac{\d s\chi^\a}{\d \epsilon^\beta}\G_{\varphi_{k+1}\ldots\varphi_l\chi_\a}
=\sqrt2 f_0\,\G_{\varphi_{k+1}\ldots\varphi_l\chi_\beta}.
\label{softbreakingterm}
\end{equation}
The fields $\varphi_{k+1}\ldots\varphi_l\chi_\beta$ have a total
dimension of $1+d$ and $R$-charge zero. Thus
$\G_{\varphi_{k+1}\ldots\varphi_l\chi_\beta}$ can receive non-zero counterterm 
contributions and it contributes to Eq.\ (\ref{dim3}) at the order
$p^{3-d}$. Its prefactor 
${\d s\chi^\alpha}/{\d \epsilon^\beta}=\sqrt2 f_0 \delta^\alpha_\beta$ 
is proportional to the soft-breaking parameter $f_0$. The situation is
similar if $n(\Delta_j)=+ n_f$ and $\varphi_k=\bar\epsilon^{\dot\a}$. In all
other cases, the term (\ref{softbreakingterm}) does not contribute in
Eq.\ (\ref{dim3}) at the considered order and to the determination 
of the counterterms.

If $n(\Delta_j)=\pm n_H$, vertex functions like $\G_{\psi_L\psi_R}$ or
$\G_{\bar\psi_L\bar\psi_R}$ that are proportional to the Higgs VEV $v$
can appear in Eq.\ (\ref{dim3}). The counterterms to these vertex functions
are then restricted by Eq.\ (\ref{dim3}). The monomials $\Delta_j$ and
$H\Delta_j$ (or $H^\dagger\Delta_j$) and the corresponding identities
are related by $R$-invariance. It is sufficient to consider only one
of them. 

If $n(\Delta_j)=0$, no symmetry-breaking parameters contribute to 
Eq.\ (\ref{dim3}). In this case, the identity (\ref{dim3}) restricts 
the counterterms to $R$-invariant and supersymmetric dimension 3 
interactions, like the $\mu$-term in the MSSM. In the model 
of Section \ref{model}, no such interactions are possible and it exists no 
$\Delta_j$ of dimension 3 that has $n(\Delta_j)=0$.
\item ${\rm dim}\Delta_j \le2$: The discussion of this case is similar
to the previous one. The different possibilities concerning the
$R$-charge are obvious, so we do not present the details.
\end{description}

\section{Evaluation and restoration of specific symmetry relations}
\label{MassRelations}

In this section we present an explicit calculation of symmetry
identities on the one-loop level and of their symmetry-restoring
counterterms. 
To be specific, we consider symmetry identities 
corresponding to a particularly direct and interesting consequence
of symmetry breakings: the appearance of non-zero masses for electrons
and selectrons and their interrelations.

In our model $R$-invariance is spontaneously broken and the electron mass is 
generated via the electron--Higgs Yukawa interaction. The tree-level 
value is given by $m_e=v y_{123}$. The selectron masses are related to 
the electron mass via softly broken supersymmetry as can be seen from 
the appearance of $m_e$ and soft-breaking parameters in the selectron 
mass matrix (\ref{selectronmass}). Furthermore, the electron mass
appears in the supersymmetry transformation
\begin{equation}
s\psi_R^\a = \ldots-\sqrt{2}\e^{\a}y_{123}\phi_{L}^{\dagger}(H^{\dagger} + v)
= \ldots-\sqrt2\e^{\a}m_e\phi_{L}^{\dagger}.
\label{transrelation}
\end{equation}
These relations are replaced by Ward and Slavnov-Taylor 
identities in higher orders
which are of the form (\ref{master}) and the strategy of Section 
\ref{sec:Renormalization} can be applied.   
In this section, all Green  
functions appearing in the Slavnov-Taylor identities are evaluated 
analytically, the one-loop breakings are calculated explicitly, and 
suitable symmetry-restoring counterterms are determined.
For the explicit calculation, we use DRED and DREG with anticommuting
$\g_5$ and adopt the 4-spinor notation introduced in Appendix \ref{app:fourspinors}.

\subsection{Electron mass relation}

The relation between the electron mass and the electron--Higgs Yukawa
coupling is expressed by the Ward identity
\begin{equation} 
\label{widirac}
0= \left. P_L\frac{\d^2 {\mathcal W}(\G)}{\delta\Psi \delta
  {\bar\Psi}}P_L \right|_{\rm \varphi = 0}=
 P_L\left( \Gamma_{\Psi \bar\Psi}  - \sqrt{2}
v \Gamma_{\Psi \bar\Psi H_2} + 3 f_0
 \Gamma_{\Psi \bar\Psi f_2} \right)P_L.
\end{equation}
Eq.\ (\ref{widirac}) describes the spontaneous breaking of $R$-invariance.
$H_2$ and $f_2$ are defined by $H_2=-\sqrt2 \i {\rm Im}H$ and 
$f_2=- \i {\rm Im}f$. This identity connects the electron-mass
contribution of the self energy $P_L\G_{\Psi\bar\Psi}P_L$ 
($=-m_e P_L+\mbox{higher orders}$) to the Yukawa coupling 
$\G_{\Psi\bar\Psi H_2}$. The vertex function 
$\Gamma_{\Psi \bar\Psi f_2}$ has no $R$-invariant and power-counting
renormalizable contribution and vanishes at tree level. At higher orders
it contains only finite loop diagrams. An explicit one-loop calculation 
shows that Eq.\ (\ref{widirac}) is valid both in DRED and DREG
(with anticommuting $\g_5$) and no symmetry-restoring counterterms are 
required. The latter result is in agreement with the validity of the full 
Ward identity ${\mathcal W}\G=0$ in DREG as discussed in Section 
\ref{secStructure}.

\subsection{Electron-selectron mass relation}

Our main 
interest in this subsection is the relation between the electron and
selectron masses since this relation is influenced by spontaneous and 
soft symmetry breaking and involves the relation (\ref{transrelation}) of the
supersymmetry transformation. The relevant Slavnov-Taylor identity can 
be obtained from
\begin{equation}
\left.
\frac{\d {\mathcal S(\Gamma)}}{
  \d \phi_L^\dagger(-p) \d \Psi(p) \d \bar \e }\right|_{\rm \varphi = 0} = 0
\end{equation}
and reads
\begin{eqnarray} 
\label{stiphys}
0 & = & \Gamma_{\Psi \bar \e Y_{\phi_L}}
  \Gamma_{\phi_L^{\dagger} \phi_L} + 
  \Gamma_{\Psi \bar \e Y_{\phi_R^{\dagger}}} \Gamma_{\phi_L^{\dagger}
  \phi_R^\dagger}
 +  \Gamma_{Y_{\bar{\hat H}} \bar \e}
  \Gamma_{\phi_L^{\dagger} \Psi \bar{\hat H}}
 +  \Gamma_{  \phi_L^{\dagger} Y_{\bar \Psi} \bar \e}
 \Gamma_{\Psi \bar\Psi} \nonumber\\
&&{} +  \frac {\partial\int {\rm d}^4x\, s \bar\Xi}{\partial \bar \e} \Big|_{\rm
  \varphi= 0} \Gamma_{\phi_L^{\dagger} \Psi \bar\Xi}.
\end{eqnarray}

Some of the Green functions in Eq.\ (\ref{stiphys}) can appear at tree level
and can receive counterterm contributions, for others the tree-level and 
counterterm contributions vanish due to $R$-invariance and power counting. 
The meaning of (\ref{stiphys}) gets
more transparent if we multiply it with the projectors $P_{L,R}$ and
distinguish those two types of contributions.
Multiplying Eq.\ (\ref{stiphys}) from left and right with $P_L$ yields
\begin{eqnarray}
0 & = & P_L\left(
\Gamma_{\Psi \bar \e Y_{\phi_L}} \Gamma_{\phi_L^{\dagger} \phi_L} + 
\Gamma_{  \phi_L^{\dagger} Y_{\bar \Psi} \bar \e} \Gamma_{\Psi \bar\Psi} + 
\sqrt2 f_0\Gamma_{\phi_L^{\dagger} \Psi \bar\Xi} \right)P_L
\nonumber\\
&&{} +\mbox{loop contributions}.
\label{stimass}
\end{eqnarray}
The ``loop contributions'' contain all Green functions that 
are given in terms of finite one-loop diagrams
and do not involve counterterms due to $R$-invariance and power counting.
This identity relates the electron and selectron self energies and the 
soft-breaking term $\propto f_0$. In lowest order it corresponds to 
the mass 
relation $(M_\phi^2){}_{11}=m_e^2+M_{11}^2$. The prefactors of the 
self energies are loop-corrected supersymmetry transformations (see
also Refs.\ \cite{Hollik:1999xh,Hollik:2001cz} for further discussion). 

These prefactors are
also restricted by the following Slavnov-Taylor 
identity which corresponds to the supersymmetry algebra,
\begin{equation}
\left.
\frac{\d {\mathcal S(\Gamma)}}{
  \d \phi_L(-p) \d \e \d \bar \e \d Y_{\phi_L}(p)}\right|_{\rm \varphi
= 0} = 0.
\label{susyalgebra}
\end{equation}
If we multiply Eq.\ (\ref{stiphys}) with $P_L$ from the left and with $P_R$ 
from the right, we obtain
\begin{eqnarray}
0 & = & P_L\left(
\Gamma_{  \phi_L^{\dagger} Y_{\bar \Psi} \bar \e}
 \Gamma_{\Psi \bar\Psi}\right) P_R + 
\mbox{loop contributions}.
\label{stisusy}
\end{eqnarray}
Written in terms of 
Weyl spinors, this identity relates in particular
$\G_{\phi_L^\dagger Y_{\psi_R}\e}$ to $\G_{\bar\psi_L\bar\psi_R}$ and,
thus, to the electron mass. This identity yields Eq.\
(\ref{transrelation}) in lowest order.

The identities obtained by multiplying Eq.\ (\ref{stiphys}) with the 
projector $P_R$ from the left are less important and are discussed later.

Both identities (\ref{stimass}) and (\ref{stisusy}) are similar to
the ones in SQED. However, Eq.\ (\ref{stimass}) contains an explicit 
soft supersymmetry breaking term $\G_{\phi_L^{\dagger} \Psi \bar\Xi}$.
Furthermore, the vertex functions in both identities receive 
contributions from the Higgs field. Because of these additional 
contributions, in particular the new $\Xi$-interactions, the breaking of 
Eqs.\ (\ref{stimass}) and (\ref{stisusy}) is modified in DREG compared 
to the SQED case, as we will see later in Eq.\ (\ref{brechung}).

In a next step, we want to gain more insight into the identity
(\ref{stiphys}) by studying the possible contributions of the breaking
term $\S(\G_{\rm reg})^{(1)}=\Delta^{(1)}$ according to the
classification of Section \ref{sec:Renormalization}. The relevant
terms in $\Delta^{(1)}$ can be expanded in the following monomials:
\begin{eqnarray}
\label{breakings}
\Delta^{(1)}|_{\phi_L^\dagger\bar\e\Psi\rm-part}
& = & \int d^4x\,\phi_L^\dagger
 \bar\e\Big(a_1 P_L + a_2 P_L\Box 
 +a_3 P_R + a_4 P_R\Box\nonumber\\&&{}
      + a_5P_L\i\g^\mu\partial_\mu+a_6P_R\i\g^\mu\partial_\mu\Big)\Psi.
\end{eqnarray}
With this notation, evaluating the r.h.s.\ of Eq.\ (\ref{stiphys}) at
the regularized level, yields
\begin{equation}
a_1 P_L - a_2 P_L p^2 +a_3 P_R - a_4 P_R p^2
        +a_5P_L\pslash+a_6P_R\pslash.
\label{breakingparameters}
\end{equation}
The dimensions and $R$-charges of the monomials corresponding to 
$a_1,\ldots,a_6$ are given in Table \ref{ta:a}.
\begin{table}[t]
\begin{center}
\begin{tabular}{|l|cccccc|}
\hline
 & $a_1$ & $a_2$ & $a_3$ & $a_4$ & $a_5$ & $a_6$ \\\hline
Dimension & 2 & 4 & 2 & 4 & 3 & 3 \\
$R$ charge & 0 & 0 & $-n_f+n_H$ & $-n_f+n_H$ & $n_H$ & $n_f$ \\
\hline
\end{tabular}
\end{center}
\label{ta:a}
\caption{Dimension and $R$-charges of the different contributions to 
the symmetry breaking corresponding to the parameters 
$a_1,\ldots,a_6$ as defined in Eq.\ (\ref{breakings})}
\end{table}
Following the results of Section \ref{sec:classification}, $a_4$ has to 
vanish since the $R$-charge has to vanish for dimension 4 monomials. 
Moreover, $a_2$ can be determined in the limit $p\to\infty$. 
In this limit, Eq.\ (\ref{stiphys}) becomes a simple relation between 
the $\phi_L$ and $\Psi$ self-energies:
\begin{eqnarray}
-a_2 P_L p^2 & = & P_L\left(\Gamma_{\Psi \bar \e Y_{\phi_L}}
  \Gamma_{\phi_L^{\dagger} \phi_L} + 
\Gamma_{  \phi_L^{\dagger} Y_{\bar \Psi} \bar \e}
 \Gamma_{\Psi \bar\Psi}\right)^{(1)}_{\rm reg}\ P_L
\nonumber\\&&{} +\mbox{subleading contributions}.
\end{eqnarray}
The soft-breaking term and the vertex functions involving the Higgs
field are negligible in this limit. 

For finite momentum $p$, we can extract from the identity (\ref{stiphys}) 
information about the remaining monomials $\propto a_{1,3,5,6}$. 
Because of $R$-symmetry, the part of the identity corresponding to 
$a_5$ contains no soft-breaking term $\propto f_0$. All other identities 
are affected by terms $\propto f_0$.

In the following, we discuss the evaluation of the identity (\ref{stiphys}) 
in DRED and DREG at one-loop order. The explicit results for the vertex 
functions can be found in Appendix \ref{app:results}. 

In DRED, the regularized loop integrals can be reduced to a
linear combination of standard scalar one- and two-point integrals,
$A_0$ and $B_0$ functions, multiplied by rational functions of 
masses and external momenta (see Ref.\ \cite{Denner:kt} for conventions). 
The scalar $A_0$ and $B_0$ integrals contain logarithms of masses and momenta. 
The logarithms have to cancel within the identity since they cannot 
contribute to local (polynomial) breaking terms as required by the Quantum 
Action Principle. These non-local parts are functions of the arguments of the 
scalar integrals. Assuming that $A_0$ and $B_0$ integrals depending 
on different, non-vanishing arguments are uniquely characterized by 
their non-local parts, we expect that the coefficients in front of the
$A_0$ and $B_0$ integrals depending on a certain set of non-vanishing 
arguments add up to zero in a symmetry identity. This is 
what happens in the identities considered above when DRED is used.
As a result no symmetry-restoring counterterms are required in this 
identity using DRED.
   
This idea is also applicable for more complicated symmetry identities. 
After reducing the tensor integrals to scalar integrals, we can choose a basis 
of scalar integrals in the sense that these one-loop integrals are uniquely 
characterized by their non-local parts, i.e.\ the logarithms and dilogarithms. 
We call two scalar integrals $\mathcal T_1$ and $\mathcal T_2$ linearly 
independent if they cannot be related by
\begin{equation}
\mathcal P_1 \mathcal T_1+\mathcal P_2 \mathcal T_2=\mathcal P_3,
\end{equation}
where $\mathcal P_{1,2,3}$ are polynomials of masses and momenta. 
These one-loop integrals then have to cancel within symmetry identities. 

The calculation of one-loop diagrams may however yield rational
functions of momenta and masses in addition to scalar integrals. 
These extra terms can originate from several sources. 
One is the reduction of tensor integrals to scalar ones, which is 
performed in $D$ dimensions for DREG as well as DRED. Certain tensor 
integrals, which have at least two Lorentz indices,  
cannot be reduced to a combination of scalar integrals alone, e.g.\ 
in the tensor reduction of $B_{11}(p^2,m_0,m_1)$ a term 
$(m_0^2+m_1^2-p^2/3)/6$ remains apart from contributions involving
scalar integrals. These 
integrals, however, do not occur in the calculation of the vertex 
functions of the r.h.s.\ of Eq.\ (\ref{stiphys}). Another source for 
polynomial terms in the calculation of symmetry identities, which
contribute in DREG but not in DRED, is the appearance of 
$\mathrm{terms} \propto(D-4)$ in the Dirac algebra, which multiply
$1/(D-4)$-parts of the scalar integrals. 
 
In the case of this particular identity, in DREG (with anticommuting
$\g_5$) we obtain a finite rational function in the limit
$D\to4$ on the r.h.s.\ of Eq. (\ref{stiphys}):
\begin{eqnarray}
\frac{\a}{4 \pi} \Big[
-\sqrt 2 (2 m_e^2 + 2 m_{\tilde \g}^2 - p^2) P_L
- 2 \sqrt 2 m_e
 m_{\tilde \g} P_R - \sqrt 2 \left( m_e P_L -  
 m_{\tilde \g} P_R \right) \pslash \Big].
\label{brechung}
\end{eqnarray}
We can directly read off the values of the coefficients
$a_1,\ldots,a_6$, in particular $a_4=0$. The breaking term is in
agreement with the Quantum Action Principle: It is a local term of
dimension 2 and, as explained in Section \ref{sec:classification}, it
is a polynomial in the mass parameters.

The appearance of the terms 
$\propto m_{\tilde{\g}}=-\tilde{M}_\lambda f_0$ shows that soft
breaking of supersymmetry induces additional violations of the
Slavnov-Taylor identity in DREG. In other words, DREG does not 
only violate supersymmetry, it even violates softly broken supersymmetry. 
Hence, in order to calculate supersymmetry-restoring counterterms, 
the soft-breaking must not be neglected.

The difference between DREG and DRED is confined to the following two 
vertex functions:
\begin{eqnarray}
\label{differencea}
\Gamma_{\Psi \bar\Psi}^{(1)}(p, -p)|_{\rm DREG-DRED} & = & 
 - \frac{\a}{4 \pi} (\pslash - 2 m_e), \\
\label{differenceb}
\Gamma_{\phi_L^{\dagger} \Psi \bar\Xi}^{(1)}(p,-p,0)|_{\rm DREG-DRED} & = &
 \frac{\a}{4 \pi} \frac{m_{\tilde\g}}{f_0} \left[\pslash 
   + 2 (m_e-m_{\tilde\g})\right]P_L.
\end{eqnarray}
\begin{figure}
\setlength{\unitlength}{1cm}
\centerline{
\begin{picture}(8.0,3.6)
\put(-5,-20.3){\includegraphics{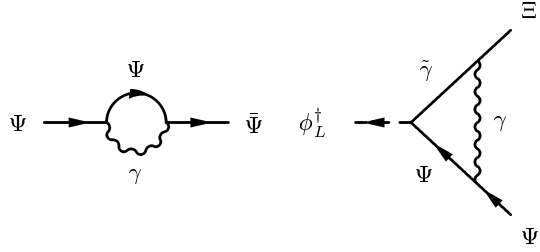}}
\end{picture}}
\label{fig:boese}
\caption{Diagrams contributing to the difference in the symmetry breaking 
(\ref{brechung}) between DREG and DRED [see Eqs.\ (\ref{differencea}) 
and (\ref{differenceb})]}
\end{figure}
The additional local contributions are generated in the diagrams shown
in Figure \ref{fig:boese}.
In particular the second diagram in Figure \ref{fig:boese} involves
a $\g\tilde{\g}\Xi$ vertex, i.e.\ an interaction of the
photon and two neutral fields. This does not correspond to a gauge
interaction but to $\eta W^\a W_\a$ in Eq.\ (\ref{Lsoft}), a type
of interaction not present in a usual supersymmetric gauge theory without 
dimensionless chiral superfields. For this reason it is noteworthy that 
this new interaction induces additional symmetry violations in DREG, 
but not in DRED.

Identities like Eq.\ (\ref{susyalgebra}) have been discussed in Refs.\ 
\cite{Hollik:1999xh,Hollik:2001cz} for SQED and SQCD. In order to 
extract the relevant restrictions on the supersymmetry transformations, 
we only need to evaluate these identities for $p\to\infty$ where 
soft breaking do not contribute. 
Hence these identities are respected by DRED and DREG.

\subsection{Parameterization of counterterms}

In this subsection,
we parameterize the counterterms that appear in
the Slavnov-Taylor identity (\ref{stiphys}). As mentioned in Section
\ref{secStructure}, the separation of the counterterms into symmetric
and symmetry-restoring ones is not unique. It becomes unique if we
impose that the symmetry-restoring counterterms do not contribute to those
Green functions on which renormalization conditions are imposed. We make
use of this possibility and denote the counterterms as follows
(the momentum $p$ is incoming into $\Psi$ and outgoing from $\phi_L^\dagger$):
\begin{eqnarray}
\G^{\rm ct}_{\phi^\dagger_L\phi_L} & = & 
[p^2-(M_\phi^2)_{11}](1+\d Z_\phi) -\d (M_\phi^2)_{11},
\\
\G^{\rm ct}_{\phi^\dagger_L\phi_R^\dagger} & = & 
-(M_\phi^2)_{12}-\d (M_\phi^2)_{12},
\\
\G^{\rm ct}_{\Psi\bar\Psi} & = & 
(\pslash - m_e)(1+\d Z_\Psi) - \d m_e,
\\
\G^{\rm ct}_{\phi_L^{\dagger} \Psi \bar{\hat H}} & = &
-P_R(y_{123} + \d y_{123}),
\\
\G^{\rm ct}_{\Psi \bar \e Y_{\phi_L}} & = &
\sqrt2 P_L(1+\d _{\psi\e Y_{\phi}}),
\\
\G^{\rm ct}_{\Psi \bar \e Y_{\phi_R^{\dagger}}} & = &
-\sqrt2 P_R(1+\d _{\psi\e Y_{\phi}}),
\\
\G^{\rm ct}_{  \phi_L^{\dagger} Y_{\bar \Psi} \bar \e}  & = &
-\sqrt2 P_L (\pslash+m_e)(1+\d _{\phi\e Y_{\psi}}^1)
- \sqrt2 P_L\, m_e\, \d_{\psi\e Y_{\psi}}^2 
- \sqrt2 P_R\, f_0\, \d u_{3\phi},
\\
\G^{\rm ct}_{Y_{\bar{\hat H}}\bar \e} & = &
\sqrt2 \g_5 \frac{y_{333}v^2}{2} - \frac{\sqrt2}{2} \g_5\, 
\left(2 f_0 \, \d u_{3H} - v^2 \d y_{333}\right),
\\
\G^{\rm ct}_{\phi_L^{\dagger} \Psi \bar\Xi} & = &
P_L \tilde{M}_{11}^2\, f_0 (1 + \d ^1_{\phi\psi\chi} ) 
+P_R \tilde{A}_{123}\, v(1+\d^2_{\phi\psi\chi}) + 
\pslash P_L\, \d ^3_{\phi\psi\chi}.
\end{eqnarray}
We have introduced the most general counterterms compatible with
Lorentz invariance, $R$-symmetry, and power counting. Moreover,
we have written down also the tree-level contributions.
The renormalization constants $\d m_e$, $\d M_\phi^2$, $\d Z_\Psi$,
$\d Z_\phi$, $\d y_{123}$, $\d u_{3\phi}$, $\d u_{3H}$ correspond to
purely symmetric counterterms, whereas the constants 
$\d_{\psi\e Y_{\phi}}$, $\d _{\phi\e Y_{\psi}}^{1,2}$, 
$\d^{1,2,3}_{\phi\psi\chi}$ correspond to a sum of symmetric and 
symmetry-restoring counterterms.

The parameters $\d u_{3\phi}$, $\d u_{3H}$ deserve some
discussion. They correspond to the $u_3$-parameters of Ref.\ 
\cite{Hollik:2000pa}. These symmetric counterterms are generated by
\begin{eqnarray}
\d u_{3\phi} \, \XS_{\Gcl}\int{\rm d}^4 x \, \bar\Xi\, P_R\, Y_{\bar\Psi}\, \phi_L^\dagger,
\quad
\d u_{3H} \, \XS_{\Gcl}\int{\rm d}^4 x \, \bar\Xi\, Y_{\bar{\hat H}}.
\end{eqnarray}
In general, $u_3$-counterterms can be written as 
$\d u_{3i} \, \XS_{\G_{\rm cl}}\int{\rm d}^4 x \, (Y_{\psi_i} \chi \phi_i)$, 
and adding these counterterms corresponds to a field renormalization of 
the form
\begin{eqnarray} \label{fren}
{\psi_i}_{\a} & \longrightarrow &{\psi_i}_{\a} - 
\d u_{3i} \chi_{\a} \phi_i.
\end{eqnarray}
The counterterms $\d u_{3i}$ do not contribute to physical amplitudes 
in the limit $\Xi = a = 0$ \cite{Hollik:2000pa}. However, they are necessary 
in order to absorb divergences of vertex functions involving external
fields. An example in
this model is the contribution
$P_R\G_{\phi_L^{\dagger} Y_{\bar \Psi} \bar \e}P_R$ since it does not
appear in $\G_{\rm cl}$, but it contains divergences at one-loop order
that have to be absorbed by $\d u_{3\phi}$.

The $\d u_{3i}$-parameters remain as undetermined constants in the symmetry
identities 
and can be fixed by imposing renormalization conditions on the 
vertex functions $P_R\G_{\phi_L^\dagger Y_{\bar\Psi}\bar\epsilon}P_R$ and
$\G_{Y_{\bar{\hat H}} \bar\e}$.

\subsection{Results for the symmetry-restoring counterterms}

In order to derive the consequences of the Slavnov-Taylor identity
(\ref{stiphys}) for the counterterms, we rewrite Eq.\ (\ref{stiphys}) in
the form of Eq.\ (\ref{master2}).
Using the parameterization of the previous section,
the breakings of Eq.\ (\ref{breakings}) yield
\begin{eqnarray}
\lefteqn{a_1 P_L - a_2 P_L p^2 +a_3 P_R - a_4 P_R p^2
        +a_5P_L\pslash+a_6P_R\pslash  = }
\nonumber\\&&{}
\sqrt2 P_L\,\d _{\psi\e Y_{\phi}}\left[p^2-(M_\phi^2)_{11}\right] +
\sqrt2 P_L\left\{\left[(p^2-(M_\phi^2)_{11}\right]\d Z_\phi-\d (M_\phi^2)_{11}\right\} 
\nonumber\\&&{}
+\sqrt2 P_R\,\d _{\psi\e Y_{\phi}}(M_\phi^2)_{12}
+\sqrt2 P_R\,\d (M_\phi^2)_{12}
\nonumber\\&&{}
-\left[\sqrt2 P_L (\pslash+m_e)\d _{\phi\e Y_{\psi}}^1 + \sqrt2 P_L\, m_e\d_{\psi\e Y_{\psi}}^2 
+\sqrt2 P_R\, f_0\, \d u_{3\phi}\right](\pslash - m_e)
\nonumber\\&&{}
-\sqrt2 P_L(\pslash+m_e)\left[(\pslash - m_e)\d Z_\Psi - \d m_e\right]
\nonumber\\&&{}
+\sqrt2 \g_5\, f_0\, \d u_{3H}\,P_R \, y_{123}
-\sqrt2 \g_5 \frac{y_{333}v^2}{2}\,P_R \, \d y_{123}
\nonumber\\&&{}
-\sqrt2\g_5\, f_0\left[P_L \, 
 \tilde{M}_{11}^2 \, f_0\, \d ^1_{\phi\psi\chi} +
P_R \,\tilde{A}_{123}\, v\,\d^2_{\phi\psi\chi} + 
\pslash P_L\, \d ^3_{\phi\psi\chi}\right].
\end{eqnarray}
The l.h.s.\ vanishes for DRED and is given by Eq.\ (\ref{brechung}) for DREG. 
The identity can be solved for $a_1,\ldots,a_6$ yielding
\begin{eqnarray}
\frac{a_1}{\sqrt2} & = & -(M_\phi^2)_{11}(\d _{\psi\e Y_{\phi}}+\d Z_\phi)
-\d (M_\phi^2)_{11}
+m_e^2(\d _{\phi\e Y_{\psi}}^1
+\d_{\psi\e Y_{\psi}}^2 
+\d Z_\Psi)
\nonumber\\&&{}
+m_e\,\d m_e + f_0^2 \tilde{M}_{11}^2\d ^1_{\phi\psi\chi},
\\
\frac{a_2}{\sqrt2} & = & -\d _{\psi\e Y_{\phi}}-\d Z_\phi+\d _{\phi\e Y_{\psi}}^1+\d Z_\Psi,
\\
\frac{a_3}{\sqrt2} & = & \d _{\psi\e Y_{\phi}}(M_\phi^2)_{12}
+\d (M_\phi^2)_{12} + m_e\, f_0\, \d u_{3\phi}
+ y_{123}\, f_0\, \d u_{3H}-\frac{y_{333}v^2}{2}\d y_{123}
\nonumber\\&&{}
-f_0\, v\,\tilde{A}_{123}\,\d^2_{\phi\psi\chi},
\label{a3}
\\
\frac{a_5}{\sqrt2} & = & - m_e\d_{\psi\e Y_{\psi}}^2 +\d m_e,\\
\frac{a_6}{\sqrt2} & = & -f_0\, \d u_{3\phi}+f_0\, \d ^3_{\phi\psi\chi}.
\label{a6}
\end{eqnarray}

Furthermore, the identity (\ref{susyalgebra}), which corresponds to the
supersymmetry algebra, is respected by DRED as well as DREG. 
Hence, it yields in addition the relation
\begin{eqnarray}
0 & = & \d _{\phi\e Y_{\psi}}^1+\d _{\psi\e Y_{\phi}}.
\label{susyalgcts}
\end{eqnarray}

We have already discussed that the identities obtained by taking 
$P_L (\ldots) P_{L,R}$ of (\ref{stiphys}) describe 
several mass relations. 
If we restrict ourselves to these
identities (they correspond to the parameters $a_1$, $a_2$, $a_5$) and
combine them with Eq.\ (\ref{susyalgcts}), we obtain
\begin{eqnarray}
\d _{\psi\e Y_{\phi}} & = & - \d _{\phi\e Y_{\psi}}^1 = \frac{\d Z_\Psi-\d
Z_\phi}{2} -\frac{a_2}{2\sqrt2},
\\
\d_{\psi\e Y_{\psi}}^2 & = & \frac{\d m_e}{m_e} -\frac{a_5}{\sqrt2 m_e},
\\
\d ^1_{\phi\psi\chi} & = &
\frac{\d Z_\Psi+\d Z_\phi}{2}+
\frac{\d (M_\phi^2)_{11}-2m_e\,\d m_e}{f_0^2\tilde{M}_{11}^2}
  +\frac{a_1}{\sqrt2 f_0^2\tilde{M}_{11}^2}.
\end{eqnarray}
All renormalization constants
appearing in these identities are fixed
except $\d Z_\phi$, $\d Z_\Psi$, $\d m_e$, $\d
(M_\phi^2)_{11}$. The latter correspond to symmetric 
counterterms, namely to the renormalization of the fields $\phi_L$,
$\Psi$ and the mass parameters $m_e$, $(M_\phi^2)_{11}$. These counterterms 
are fixed by renormalization conditions to the electron and selectron self
energies. 

The results of Eqs.\ (\ref{a3}) and (\ref{a6}) corresponding to $a_3$
and $a_6$ can be solved similarly in 
order to obtain information on the counterterms $\d^2_{\phi\psi\chi}$ and 
$\d ^3_{\phi\psi\chi}$. The solutions, however, involve many 
renormalization constants corresponding to symmetric counterterms, in 
particular to the $u_3$-parameters. 

Although unphysical $u_3$-parameters appear in the identity (\ref{a3}), 
it is a relevant relation between the off-diagonal selectron mass
$(M_\phi^2)_{12}$, the $\phi^\dagger_L\Psi{\bar{\hat{H}}}$ Yukawa coupling,
and the 
$\Xi$-interaction $P_R\G_{\phi_L^\dagger\Psi\bar\Xi}P_R$. The
latter interaction is particularly interesting since it originates
from the term $\eta\Phi_1\Phi_2\Phi_3$ in $\L_{\rm soft}$ 
(see Appendix \ref{lagrangian}). Such an interaction is not present in  
usual renormalizable models where no dimensionless chiral superfields 
are present. In spite of this new interaction, the breaking term $a_3$ 
in Eq.\ (\ref{a3}) vanishes in DRED, but as expected it is non-zero in DREG.

Finally, we give the symmetry-restoring counterterm
contributions to the Green functions in Eq.\ (\ref{stiphys}):
\begin{eqnarray}
\G^{\rm ct,restore}_{\phi^\dagger_L\phi_L} & = & 
0,
\\
\G^{\rm ct,restore}_{\phi^\dagger_L\phi_R^\dagger} & = & 
0,
\\
\G^{\rm ct,restore}_{\Psi\bar\Psi} & = & 
0,
\\
\G^{\rm ct,restore}_{\phi_L^{\dagger} \Psi \bar{\hat H}} & = &
0,
\\
\G^{\rm ct,restore}_{\Psi \bar \e Y_{\phi_L}} & = &
- P_L\,\frac{a_2}{2},
\\
\G^{\rm ct,restore}_{\Psi \bar \e Y_{\phi_R^{\dagger}}} & = &
 P_R\,\frac{a_2}{2},
\\
\G^{\rm ct,restore}_{  \phi_L^{\dagger} Y_{\bar \Psi} \bar \e}  & = &
-  P_L (\pslash+m_e)\frac{a_2}{2}
+  P_L\, a_5 ,
\\
\G^{\rm ct,restore}_{Y_{\bar{\hat H}}\bar \e} & = &
0,
\\
\G^{\rm ct,restore}_{\phi_L^{\dagger} \Psi \bar\Xi} & = &
\frac{1}{\sqrt2 f_0}(P_L \,{a_1}
- P_R{a_3} + 
\pslash P_L\, a_6).
\end{eqnarray}

\section{Summary and Conclusion}

In this article we have considered a model with characteristic
features of the MSSM, i.e. soft supersymmetry breaking and spontaneous
symmetry breaking.

We have evaluated two important symmetry identities corresponding to mass
relations for arbitrary momenta of the external fields in
DREG as well as DRED. The Ward identity relating the electron mass to
the Yukawa coupling is preserved both in DRED and in DREG. The
Slavnov-Taylor identity relating the electron and selectron self
energies is preserved in DRED, but violated in DREG. 

We have shown that all appearing symmetry-restoring counterterms are 
uniquely fixed when the Slavnov-Taylor identity relating electron and 
selectron self energies is combined with an identity corresponding to 
the supersymmetry algebra. 

Taking a closer look how the Slavnov-Taylor identity is violated in 
DREG reveals that the coefficients of all possible breaking terms 
[see Eq.\ (\ref{breakingparameters})] are all non-vanishing. Moreover, 
in all cases where soft breaking is involved, soft breaking does actually 
affect the symmetry-restoring counterterms. 
In particular, the spurion field $\eta$ and its spinor component $\Xi$
give rise to $\eta W^\a W_\a$ and $\gamma\tilde{\gamma}\Xi$
interactions, which contribute to the violations of the Slavnov-Taylor identity 
in DREG, but not in DRED.

These results have important implications for the MSSM, where e.g.\
the left-handed stop and sbottom masses $m_{\tilde{t}_L}$,
$m_{\tilde{b}_L}$ are related to the $\Xi$-interactions
$\tilde{t}_L^\dagger t\bar\Xi$ and $\tilde{b}_L^\dagger b\bar\Xi$.
These $\Xi$-interactions can be eliminated using SU(2) invariance 
and an identity between $m_{\tilde{t}_L}^2-m_{\tilde{b}_L}^2$
and $m_t^2-m_b^2$ can be derived. Assuming that the results of the 
model considered here hold also for the MSSM, the identity for
$m_{\tilde{t}_L}^2-m_{\tilde{b}_L}^2$, which is an important prediction of
the MSSM, is fulfilled in DRED, but violated in DREG. Indeed
the full (electroweak and strong) one-loop results of this relation 
evaluated in DRED and DREG without symmetry-restoring counterterms
differ by finite contributions which are of the order of the full 
electroweak corrections \cite{Heidi}.

Furthermore, several simplifications in the determination of 
symmetry-restoring counterterms have been 
discussed. The breaking $\Delta$ of a symmetry 
identity is a polynomial not only in the momenta, but also in the 
mass parameters of the model. Symmetry identities that correspond to 
$\mathrm{dimension} = 4$ breakings can be evaluated in
the limit $p\to \infty$, $m\to0$. In symmetry identities
corresponding to $\mathrm{dimension} < 4$ breakings, masses and 
soft supersymmetry breaking cannot be neglected. However, in order to 
compute symmetry violations and symmetry-restoring counterterms, it is 
sufficient to evaluate the identities in a Taylor expansion in the masses. 
This is an important simplification in the restoration of symmetry identities
in the MSSM where the structure of masses and mixings are rather complicated.

\begin{appendix}
\section*{Appendix}

\section{Lagrangian}
\label{lagrangian}

The Lagrangian $\L$ is obtained from $\L_{\rm susy}+\L_{\rm soft}$ by
choosing the Wess-Zumino gauge and eliminating the auxiliary fields.
We specify the Lagrangian of the model in terms of mass
eigenstates. The mass eigenstates and their eigenvalues are
\begin{equation}
\begin{array}[b]{ll}
\phi_1 = \frac{1}{\sqrt{2}}(\phi_L + \phi_R^{\dagger}), & M_1^2
= (y^{123})^2 v^2 + \tilde M_{11}^2 f_0^2 + \frac 12 y^{123} y^{333} v^2
  + 6 f_0 \tilde A_{123} v, \\
\phi_2 = \frac{1}{\sqrt{2}}(\phi_R - \phi_L^{\dagger}), & 
M_2^2 = (y^{123})^2 v^2 + \tilde M_{11}^2 f_0^2 - \frac 12 y^{123} y^{333} v^2
  - 6 f_0 \tilde A_{123} v
\end{array}
\end{equation}
as well as
\begin{equation}
\begin{array}[b]{ll}
H_1 = \frac{1}{\sqrt 2} (H + H^{\dagger}) = H_1^{\dagger}, & \frac 12 M_{H_1}^2 = 
\frac 34 (y^{333})^2 v^2 + \frac 12 \tilde M_{33}^2 f_0^2 + 3 f_0 \tilde A_{333} v, \\
H_2 = \frac{1}{\sqrt 2} (- H + H^{\dagger}) = - H_2^{\dagger}, & \frac
12 M_{H_2}^2
= \frac 14 (y^{333})^2 v^2 + \frac 12 \tilde M_{33}^2 f_0^2 - 3 f_0 \tilde A_{333} v. 
\end{array}
\end{equation}
The Lagrangian of the model in the limit $a = \Xi = f = 0$ reads
\begin{eqnarray}
\lefteqn{{\mathcal L}|_{a=\Xi=f=0} =}\nonumber\\
&&{} -\frac 14 F_{\m\n}F^{\m\n} + \frac 12 \bar{\tilde\gamma} \i \g^{\m}
\partial_{\mu} \tilde\g + \frac 12 f_0 \ \tilde M_{\l} \bar{\tilde\g} \tilde\g\nonumber\\
&&{}  + \bar\Psi\i\g^{\m} D_{\m}\Psi - y^{123} v \bar\Psi\Psi \nonumber\\
&&{}  + \frac 12 \bar{\hat H} \i \g^{\m}\partial_{\m} \hat H - \frac 12 y^{333} v \bar{\hat H} \hat H \nonumber\\
&&{}  + |D_{\m}\phi_1|^2 +
  |D_{\m}\phi_2|^2 \nonumber\\
&&{}  - \left[(y^{123})^2 v^2 + \tilde M_{11}^2 f_0^2 + 6 f_0 \
  \tilde A_{123} v + \frac 12 y^{123} y^{333} v^2 \right] |\phi_1|^2 \nonumber\\
&&{}  - \left[(y^{123})^2 v^2 + \tilde M_{11}^2 f_0^2 - 6 f_0 \
  \tilde A_{123} v - \frac 12 y^{123} y^{333} v^2 \right] |\phi_2|^2 \nonumber\\
&&{}  + \frac 12 (\partial^{\m} H_1^{\dagger} )(\partial_{\m} H_1) + \frac 12 (\partial^{\m} H_2^{\dagger} )(\partial_{\m} H_2) \nonumber\\
&&{}  - \frac 12 \left[\frac 32 (y^{333})^2 v^2 + \tilde M_{33} f_0^2 +
  6vf_0 \tilde A_{333}\right] |H_1|^2 \nonumber\\
&&{}  - \frac 12 \left[\frac 12 (y^{333})^2 v^2 + \tilde M_{33} f_0^2 -
  6vf_0 \tilde A_{333}\right] |H_2|^2\nonumber\\
&&{}  - e Q_L \left( \bar\Psi\g_5\tilde\g \phi_1 - \bar\Psi \tilde\g \phi_2^{\dagger} - \phi_1^{\dagger}\bar{\tilde \g}\g_5\Psi -
\phi_2 \bar{\tilde\g} \Psi \right) \nonumber\\
&&{}  - \frac{1}{\sqrt 2} y^{123} \left[ H_1 (\bar\Psi \Psi) +
  H_2(\bar\Psi\g_5\Psi) - \phi_2 (\bar{\hat H}\g_5\Psi) + \phi_1^{\dagger}(\bar{\hat H}\Psi)
    + \phi_1 (\bar\Psi\hat H) + \phi_2^{\dagger} (\bar\Psi\g_5\hat H)\right] \nonumber\\
&&{}  - \frac{1}{2\sqrt 2} y^{333} [H_1 (\bar{\hat H}\hat H) + H_2
  (\bar{\hat H}\g_5\hat H)]\nonumber\\
&&{}  - \frac{1}{\sqrt 2} \left[2 (y^{123})^2 v + y^{123}
  y^{333} v + 6 f_0 \tilde A_{123}\right] H_1 |\phi_1|^2 \nonumber\\
&&{}  - \frac{1}{\sqrt 2} \left[2 (y^{123})^2 v - y^{123}
  y^{333} v - 6 f_0 \tilde A_{123}\right] H_1 |\phi_2|^2 \nonumber\\
&&{}  - \frac{1}{\sqrt 2} \left(y^{123}
  y^{333} v - 6 f_0 \tilde A_{123}\right) H_2 \phi_1 \phi_2 \nonumber\\
&&{}  - \frac{1}{\sqrt 2} \left(- y^{123}
  y^{333} v + 6 f_0 \tilde A_{123}\right) H_2 \phi_1^{\dagger} \phi_2^{\dagger} \nonumber\\
&&{}  - \frac{1}{\sqrt 2} \left[\frac 12 (y^{333})^2 v + f_0 \tilde A_{333}\right]H_1
|H_1|^2 - \frac{1}{\sqrt 2} \left[\frac 12 (y^{333})^2 v - 3 f_0 \tilde A_{333}\right]H_1 |H_2|^2\nonumber\\
\allowdisplaybreaks[3]
&&{}  - \left[\frac 12 e^2 - \frac 14 (y^{123})^2 \right]
  \phi_1\phi_1\phi_2\phi_2 - \left[\frac 12 e^2 - \frac 14 (y^{123})^2\right]
  \phi_1^{\dagger}\phi_1^{\dagger}\phi_2^{\dagger}\phi_2^{\dagger} \nonumber\\
&&{}  - e^2 |\phi_1|^2|\phi_2|^2 - \frac 14 (y^{123})^2 \left(|\phi_1|^4 +
|\phi_2|^4\right) \nonumber\\
&&{}  + \left[ - \frac 12 (y^{123})^2 - \frac 14 y^{123} y^{333} \right]
\left(|\phi_1|^2 |H_1|^2 + |\phi_2|^2 |H_2|^2 \right) \nonumber\\
&&{}   + \left[ - \frac 12 (y^{123})^2 + \frac 14 y^{123} y^{333} \right]
\left(|\phi_1|^2 |H_2|^2 + |\phi_2|^2 |H_1|^2 \right) \nonumber\\
&&{}  - \frac 12 y^{123} y^{333}H_1 H_2
    (\phi_1\phi_2 - \phi_1^{\dagger}\phi_2^{\dagger})\nonumber\\
&&{}  - \frac{1}{16} (y^{333})^2 \left(|H_1|^4 + 2 |H_1|^2 |H_2|^2 + |H_2|^4 \right) \nonumber\\ 
&&{}  - \left[\frac{1}{\sqrt 2} v^3 (y^{333})^2 + 3 \sqrt 2 v^2 f_0 \tilde
  A_{333} + \sqrt 2 \tilde M_{33}^2 f_0^2 v \right] H_1 \nonumber\\
&&{}  - \frac 14 (y^{333})^2 v^4 - 2 f_0 \tilde A_{333} v^3 - \tilde
M_{33}^2 f_0^2 v^2.
\label{lagr}
\end{eqnarray}
In Eq.\ (\ref{lagr}) we have transformed the scalar fields to their mass
eigenstates, used four-component spinors for brevity, and introduced the
covariant derivative $D_\m$ and the electromagnetic field strength tensor $F_{\m\n}$: 
\begin{eqnarray}
D_{\m} & = & \partial_\m + \i e Q A_{\m}, \nonumber\\
F_{\m\n} & = & \partial_\m A_\n - \partial_\n A_\m.
\end{eqnarray}
All parameters in the Lagrangian ${\cal L}|_{a=\Xi=f=0}$ are real. 
For the evaluation of Slavnov-Taylor identities, it is necessary to know
the parts of the Lagrangian 
that are linear in the fields $\Xi$ and $f$. These parts of the 
Lagrangian are given by
\begin{eqnarray}
\lefteqn{\L|_{\Xi-\rm part} =}\nonumber\\
&&{} \frac{1}{\sqrt 2} \left( \tilde M_{11}^2 f_0 + 6 \tilde A_{123} v \right)
   \left[\phi_1 \bar\Psi \Xi + \phi_1^{\dagger} \bar\Xi \Psi\right] \nonumber\\
&&{} + \frac{1}{\sqrt 2} \left( \tilde M_{11}^2 f_0 - 6 \tilde A_{123} v \right)
   \left[\phi_2 \bar\Xi (P_R - P_L) \Psi + \phi_2^{\dagger} \bar\Psi (P_L -
   P_R) \Xi  \right] \nonumber\\
&&{} + \frac{1}{\sqrt 2} \left( \tilde M_{33}^2 f_0 +  6 \tilde A_{333}
   v \right) \, \bar\Xi \hat H H_1 \nonumber\\
&&{} + \frac{1}{\sqrt 2} \left( \tilde M_{33}^2 f_0 -  6 \tilde A_{333}
   v \right) \, \bar\Xi (P_L - P_R) \hat H H_2 \nonumber\\
&&{} + \frac{\i}{2 \sqrt 2} \tilde M_{\l} \bar\Xi (P_R - P_L)
   \g^{\n}\g^{\m} \tilde \g F_{\m\n} \nonumber\\
&&{} + 3 \tilde A_{123} \Big[ \Big. (\bar\Xi \Psi) \left(\phi_1^{\dagger}
   H_1 - \phi_2 H_2 \right) + (\bar\Xi (P_R - P_L) \Psi) \left(\phi_1^{\dagger}
   H_2 - \phi_2 H_1 \right) \nonumber\\
&&{} \qquad + (\bar\Psi \Xi) \left(\phi_1 H_1 - \phi_2^{\dagger} H_2
   \right) + (\bar\Psi (P_R - P_L) \Xi) \left(\phi_1 H_2 - \phi_2^{\dagger} H_1
   \right) \nonumber\\
&&{} \qquad + \bar\Xi \hat H \left(
   \phi_1^{\dagger} \phi_1 - \phi_2^{\dagger} \phi_2 \right) + \bar\Xi (P_L - P_R) \hat H \left(
   \phi_1^{\dagger} \phi_2^{\dagger} - \phi_1 \phi_2 \right) \Big. \Big] \nonumber\\
&&{} - \frac{1}{\sqrt 2} e \tilde M_{\l} (\bar\Xi \tilde\g) (\phi_1
   \phi_2 + \phi_1^{\dagger} \phi_2^{\dagger}) \nonumber\\
&&{} +{\mathcal O} (a, a^{\dagger}),
\end{eqnarray}  
and
\begin{eqnarray}
\lefteqn{\L|_{f-\rm part}=}\nonumber\\
&&{} - \left( 2 f_0 \tilde M_{11}^2 + 6
  \tilde A_{123} v \right) |\phi_1|^2 f_1 - \left( 2 f_0 \tilde M_{11}^2 - 6
  \tilde A_{123} v \right) |\phi_2|^2 f_1\nonumber\\
&&{} - \frac 12 \left( 2 f_0 \tilde M_{33}^2 + 6
  \tilde A_{333} v \right) |H_1|^2 f_1 - \frac 12 \left( 2
f_0 \tilde M_{33}^2 - 6
  \tilde A_{333} v \right)|H_2|^2 f_1 \nonumber\\
&&{} + 6
\tilde A_{123} v \phi_1 \phi_2 f_2 - 6 \tilde A_{123} v
\phi_1^{\dagger} \phi_2^{\dagger} f_2 - 6 \tilde A_{333} v H_1 H_2 f_2 \nonumber\\
&&{} + \frac 12 \tilde M_{\l} f_1 \bar{\tilde\g} \tilde\g - \frac 12 \tilde
  M_{\l} f_2 \bar{\tilde\g} \g_5 \tilde\g \nonumber\\
&&{} +{\mathcal O} (a, a^{\dagger}),
\end{eqnarray}
where we used Eq.\ (\ref{Chi}) and
\begin{equation} 
\label{xi}
f_{1,2}  =  \frac 12 \left(\pm f + f^{\dagger} \right).
\end{equation}

The gauge fixing and ghost terms are combined in a BRS variation
\begin{eqnarray}
\label{gaugefix}
{\cal L}_{\rm fix, gh} & = & s\left[\bar c \left(\partial^{\m} A_{\m} +
\frac{\xi}{2}B\right)\right] \nonumber\\
& = & B \partial^{\m}A_{\m} + \frac{\xi}{2}B^2
- \bar c \square c - \bar c \partial^{\m} (\i\e\s_{\m}\bar\l - \i\l\s
_{\m}\bar\e) + \xi\i\e\s^{\n}\bar\e(\partial_{\n}\bar c) \bar c.
\end{eqnarray}
When the auxiliary field $B$ is eliminated, Eq.\ (\ref{gaugefix}) yields 
the usual gauge-fixing term $-(\partial_{\m} A^{\m})^2/(2 \xi)$ 
and ghost terms, in particular the $\bar{c}\epsilon^\a\lambda_\a$ vertex 
due to the supersymmetry breaking of the gauge fixing.

The BRS transformations given in Appendix \ref{brs} that are non linear 
in the dynamical fields are coupled to external sources,
\begin{eqnarray} 
\label{Gext}
{\cal L}_{\rm ext} & = & Y_{\l}^{\a} s \l_{\a} +
  Y_{\bar\l\dot\a} s \bar\l^{\dot\a} \Big. \nonumber\\
&&{} + Y_{\phi_L} s \phi_L + Y_{\phi_L^{\dagger}} s \phi_L^{\dagger} +
  Y_{\psi_L}^{\a} s \psi_{L\a} + Y_{\bar\psi_L \dot\a} s
  \bar\psi_L^{\dot\a} + (_{L \rightarrow R}) \nonumber\\
&&{} + Y_{\tilde H}^{\a} s \tilde
  H _{\a} + Y_{\bar{\tilde H} \dot\a} s \bar{\tilde H}^{\dot\a}
  \nonumber\\
&&{} + \Big. \frac 12 (Y_{\l}^\a\e_\a +
  Y_{\bar\l \dot\a}\bar\e^{\dot\a})^2 - 
  2(Y_{\tilde H}^\a\e_\a)(\bar\e_{\dot\a} Y_{\bar{\tilde H}}^{\dot\a}) 
\nonumber\\
&&{} - 2(Y_{\psi_L}^\a\e_\a)(\bar\e_{\dot\a} Y_{\bar\psi_L}^{\dot\a}) -
  2(Y_{\psi_R}^\a \e_\a)(\bar\e_{\dot\a} Y_{\bar\psi_R}^{\dot\a}). 
\end{eqnarray}

The classical action is given by
\begin{equation}
\G_{\rm cl} = \int{\rm d}^4 x \, \left({\mathcal L} + 
{\mathcal L}_{\rm fix,gh} +{\mathcal L}_{\rm ext}\right)
\end{equation}
and satisfies the Slavnov-Taylor identity $S(\G_{\rm cl})=0$.

\section{BRS transformations}
\label{brs}

The BRS transformations of the fields of the model have the following 
explicit form: 
\begin{eqnarray}
sA_{\m} & = & \partial_{\m} c + \i\e\s_{\m}\bar\l - \i\l\s_{\m}\bar\e -
\i\w^{\n}\partial_{\n}A_{\m}, \\
s\l^{\a} & = & \frac{\i}{2}(\e\s^{\r\s})^{\a}F_{\r\s} -
\i\e^{\a}eQ_{L}(|\phi_{L}|^2 - |\phi_{R}|^2) -
\i\w^{\n}\partial_{\n}\l^{\a}, \\
s\bar\l_{\dot\a} & = & \frac{-\i}{2}({\bar\e\bar\s}^{\r\s})_{\dot\a}F_{\r\s} -
\i\e_{\dot\a}eQ_{L}(|\phi_{L}|^2 - |\phi_{R}|^2) -
\i\w^{\n}\partial_{\n}\bar\l_{\dot\a}, \\
s\phi_{L} & = & -\i eQ_{L}c\phi_{L} + \sqrt{2}\e\psi_{L} -
\i\w^{\n}\partial_{\n}\phi_{L}, \\
s\phi_{L}^{\dagger} & = & +\i eQ_{L}c\phi_{L}^{\dagger} + \sqrt{2}\bar\psi_{L}\bar\e -
\i\w^{\n}\partial_{\n}\phi_{L}^{\dagger}, \\
s\psi_{L}^{\a} & = & -\i eQ_{L}c\psi_{L}^{\a} -
\sqrt{2}\e^{\a}y^{123}\phi_{R}^{\dagger}(H^{\dagger} + v) 
 - \sqrt{2}\i(\bar\e\bar\s^{\m})^{\a}D_{\m}\phi_{L} \nonumber\\
&&{} - \i\w^{\n}\partial_{\n}\psi_{L}^{\a}, \\
s\bar\psi_{L\dot\a} & = & +\i eQ_{L}c\bar\psi_{L\dot\a} +
\sqrt{2}\bar\e_{\dot\a}y^{123}\phi_{R} (H + v) +
\sqrt{2}\i(\e\s^{\m})_{\dot\a}(D_{\m}\phi_{L})^{\dagger} \nonumber\\
&&{} - \i\w^{\n}\partial_{\n}\bar\psi_{L\dot\a}, \\
s\phi_{R} & = & -\i eQ_{R}c\phi_{R} + \sqrt{2}\e\psi_{R} -
\i\w^{\n}\partial_{\n}\phi_{R}, \\
s\phi_{R}^{\dagger} & = & +\i eQ_{R}c\phi_{R}^{\dagger} + \sqrt{2}\bar\psi_{R}\bar\e -
\i\w^{\n}\partial_{\n}\phi_{R}^{\dagger}, \\
s\psi_{R}^{\a} & = & -\i eQ_{R}c\psi_{R}^{\a} -
\sqrt{2}\e^{\a}y^{123}\phi_{L}^{\dagger}(H^{\dagger} + v) 
 - \sqrt{2}\i(\bar\e\bar\s^{\m})^{\a}D_{\m}\phi_{R} \nonumber\\
&&{} - \i\w^{\n}\partial_{\n}\psi_{R}^{\a}, \\
s\bar\psi_{R\dot\a} & = & +\i eQ_{R}c\bar\psi_{R\dot\a} +
\sqrt{2}\bar\e_{\dot\a}y^{123}\phi_{L} (H + v) +
\sqrt{2}\i(\e\s^{\m})_{\dot\a}(D_{\m}\phi_{R})^{\dagger} \nonumber\\
&&{} - \i\w^{\n}\partial_{\n}\bar\psi_{R\dot\a}, \\
s(H + v) & = & \sqrt{2} \e \tilde H - \i \w^{\n}\partial_{\n} H,\\
s(H^{\dagger} + v) & = & \sqrt{2} \bar{\tilde H}\bar\e - \i
\w^{\n}\partial_{\n} H^{\dagger},\\
s\tilde H^{\a} & = & - \sqrt{2}\e^{\a}\left[y^{123}
  \phi_L^{\dagger} \phi_R^{\dagger} + \frac 12 y^{333} (H^{\dagger} + v)
  (H^{\dagger} + v) \right] \nonumber\\
&&{} - \sqrt{2} \i (\bar\e\bar\s^{\m})^{\a}\partial_{\m} H -
\i\w^{\n}\partial_{\n}\tilde H^{\a}, \\ 
s\rwu{\tilde H}{\a} & = & + \sqrt{2}\rwu{\e}{\a}\left[y^{123}
  \phi_L \phi_R + \frac 12 y^{333} (H + v) (H + v)\right] \nonumber\\
&&{} + \sqrt{2} \i
(\e\s^{\m})_{\dot\a}\partial_{\m} H^{\dagger} -
\i\w^{\n}\partial_{\n}\rwu{\tilde H}{\a}, \\ 
sc & = & 2\i\e\s^{\n}\bar\e A_{\n} - \i \w^{\n}\partial_{\n}c, \\
s\e^{\a} & = & 0, \\
s\bar\e^{\dot\a} & = & 0, \\
s\w^{\n} & = & 2\e\s^{\n}\bar\e, \\
s\bar c & = & B - \i\w^{\n}\partial_{\n}\bar c, \\
sB & = & 2\i\e\s^{\n}\bar\e\partial_{\n}\bar c - \i \w^{\n}\partial_{\n} B.
\end{eqnarray}
The ghosts $c$, $\e_\a$, $\bar\e^{\dot\a}$, and $\w^{\n}$ correspond
to gauge, supersymmetry transformations, and translations. The antighost 
$\bar c$ and the auxiliary field $B$ are needed for gauge fixing. 

The BRS transformations of the fields of the external spurion multiplet
are given by
\begin{eqnarray}
sa & = & \sqrt 2 \e \chi - \i\w^{\n}\partial_{\n} a, \\
sa^{\dagger} & = & \sqrt 2 \bar\chi\bar\e - \i\w^{\n}\partial_{\n}
a^{\dagger}, \\
s\chi^{\a} & = & \sqrt 2 \e^{\a} \hat f - \sqrt 2 \i
(\bar\e\bar\s^{\m})^{\a}\partial_{\m} a - \i
\w^{\n}\partial_{\n}\chi^{\a}, \\
s\rwu{\chi}{\a} & = & - \sqrt 2 \rwu{\e}{\a} \hat f^{\dagger} + \sqrt 2 \i
(\e\s^{\m})_{\dot\a}\partial_{\m} a^{\dagger} - \i
\w^{\n}\partial_{\n}\rwu{\chi}{\a}, \\
sf & = & \sqrt 2 \i\bar\e\bar\s^{\m}\partial_{\m}\chi - \i\w^{\n}\partial_{\n} f, \\
sf^{\dagger} & = & - \sqrt 2 \i\partial_{\m}\bar\chi\bar\s^{\m}\e - \i\w^{\n}\partial_{\n} f^{\dagger}. 
\end{eqnarray}

\section{4-spinor notation}
\label{app:fourspinors}

We use the following conventions for derivatives with respect to Weyl spinors:
\begin{equation}
\frac{\d}{\d\psi_{\a}} \psi_{\b} = -{\d^{\a}}_{\b}, \quad  
\frac{\d}{\d\psi^{\a}} \psi^{\b} ={\d_{\a}}^{\b}, \quad 
\frac{\d}{\d\rwo{\psi}{\a}} \rwo{\psi}{\b} = -{\d_{\dot\a}}^{\dot\b}, 
\quad  
\frac{\d}{\d\rwu{\psi}{\a}} \rwo{\psi}{\b} ={\d^{\dot\a}}_{\dot\b}.
\end{equation}
The 4-spinors and derivatives with respect to them are defined in such 
a way that \ $(\d / \d\Psi) \, \Psi = 1$ and 
$(\d / \d\bar\Psi) \, \bar\Psi = 1$.
\begin{itemize}
\item Electron:
\begin{align}
\Psi & = \left( \begin{array}{c}{\psi_L}_{\a} \\ \rwo{\psi_R}{\a} \end
 {array} \right) , &
\frac{\d}{\d\Psi} & = \left( \begin{array}{cc} \displaystyle - \frac{\d}{\d
     {\psi_L}_{\a}} & \displaystyle - \frac{\d}{\d \rwo{\psi_R}{\a}}
  \end{array} \right), \\
\bar \Psi & = \left( \begin{array}{cc}{\psi_R}^{\a} & \rwu{\psi_L}{\a}
  \end{array} \right) , &
\frac{\d}{\d\bar\Psi} & = \left( \begin{array}{c} \displaystyle \frac{\d}{\d \psi_R^{\a}}. \\
    \displaystyle \frac{\d}{\d \rwu{\psi_L}{\a}} \end{array} \right).
\end{align}
Sources for the BRS transformations of the electrons:
\begin{align}
Y_{\Psi} & = \left( \begin{array}{cc} Y_{\psi_L}^{\a} &
   {Y_{\bar\psi_R}}_{\dot\a} \end{array} \right), & 
\frac{\d}{\d Y_{\Psi}} & = \left( \begin{array}{c} \displaystyle \frac{\d}{\d
    Y_{\psi_L}^{\a}} \\ \displaystyle \frac{\d}{\d{Y_{\bar\psi_R}}_{\dot\a}} \end{array} \right),\\ 
Y_{\bar\Psi} & = \left( \begin{array}{c} -{Y_{\psi_R}}_{\a} \\ - Y_{\bar
    \psi_L}^{\dot \a} \end{array} \right), & 
\frac{\d}{\d Y_{\bar\Psi}} & = \left( \begin{array}{cc} \displaystyle
    \frac{\d}{\d{Y_{\psi_R}}_{\a}} & \displaystyle \frac{\d}{\d Y_{\bar
    \psi_L}^{\dot \a}} \end{array} \right).
\end{align}
\item Higgsino:
\begin{align}
\hat H & = \left( \begin{array}{c}{\tilde H_{\a}} \\ \rwo{\tilde H}{\a}
  \end{array} \right) , &
\frac{\d}{\d\hat H} & = \left( \begin{array}{cc} \displaystyle - \frac{\d}{\d
      \tilde H_{\a}} & \displaystyle - \frac{\d}{\d \rwo{\tilde H}{\a}}
  \end{array} \right), \\
Y_{\hat H} & = \left( \begin{array}{cc} Y_{\tilde H}^{\a} &
   {Y_{\bar{\tilde H}}}_{\dot\a} \end{array} \right) , & 
\frac{\d}{\d Y_{\hat H}} & = \left( \begin{array}{c} \displaystyle \frac
   {\d}{\d Y_{\tilde H}^{\a}} \\ \displaystyle \frac{\d}{\d
   {Y_{\bar{\tilde H}}}_{\dot\a}} \end{array} \right).  
\end{align}
\item Photino:
\begin{align}
\tilde\g & = \left( \begin{array}{c}- \i \l_{\a} \\ \i \rwo{\l}{\a}
  \end{array} \right), &
\frac{\d}{\d\tilde\g} & = \left( \begin{array}{cc} \displaystyle - \i
    \frac{\d}{\d\l_{\a}} & \displaystyle \i \frac{\d}{\d \rwo{\l}{\a}} 
  \end{array} \right), \\
Y_{\tilde\g} & = \left( \begin{array}{cc} \i Y_{\l}^{\a} & - \i 
   {Y_{\bar\l}}_{\dot\a} \end{array} \right), & 
\frac{\d}{\d Y_{\tilde\g}} & = \left( \begin{array}{c} \displaystyle - \i
    \frac{\d}{\d Y_{\l}^{\a}} \\ \displaystyle \i \frac{\d}{\d
   {Y_{\bar\l}}_{\dot\a}} \end{array} \right).  
\end{align}
\item Spinor component of the spurion superfield:
\begin{align}
\Xi & = \left( \begin{array}{c} \chi_{\a} \\ \rwo{\chi}{\a}
  \end{array} \right), &
\frac{\d}{\d\Xi} & = \left( \begin{array}{cc} \displaystyle - \frac{\d}{\d
      \chi_{\a}} & \displaystyle - \frac{\d}{\d \rwo{\chi}{\a}}
  \end{array} \right). 
\end{align}
\item Supersymmetry ghost:
\begin{align}
\e & = \left( \begin{array}{c} \e_{\a} \\ \rwo{\e}{\a}
  \end{array} \right), &
\frac{\partial}{\partial \e} & = \left( \begin{array}{cc} \displaystyle - \frac{\partial}{\partial
      \e_{\a}} & \displaystyle - \frac{\partial}{\partial \rwo{\e}{\a}}
  \end{array} \right). 
\end{align}
\end{itemize}
In this notation the Slavnov-Taylor operator is given by
\begin{eqnarray}
\S(\F) & = & \S_0(\F) + \S_{\rm soft}(\F)\nonumber\\
& = & \int{\rm d}^4x \, \bigg[ \bigg. sA^{\m} \frac{\d\F}{\d A^{\m}} + sc
  \frac{\d\F}{\d c} + s\bar c \frac{\d\F}{\d\bar c} + sB \frac{\d\F}{\d
  B} + \left( \frac{\d\F}{\d Y_{\tilde\gamma}} \right)^T \left(
  \frac{\d\F}{\d\tilde\gamma} \right)^T \nonumber\\
& & \quad + \left( \frac{\d\F}{\d Y_{\Psi}} \right)^T \left( \frac{\d\F}{\d\Psi}\right)^T +
  \frac{\d\F}{\d Y_{\bar\Psi}}\frac{\d\F}{\d\bar\Psi} \nonumber\\
& & \quad + \frac{\d\F}{\d Y_{\phi_{L}}}\frac{\d\F}{\d \phi_{L}} + 
  \frac{\d\F}{\d Y_{\phi_{L}^{\dagger}}}\frac{\d\F}{\d
  \phi_{L}^{\dagger}} + \frac{\d\F}{\d Y_{\phi_R}}\frac{\d\F}{\d \phi_R} + 
  \frac{\d\F}{\d Y_{\phi_R^{\dagger}}}\frac{\d\F}{\d
  \phi_R^{\dagger}} \nonumber\\
& & \quad + s(H + v)\frac{\d\F}{\d H} + s(H^{\dagger} + v) \frac{\d\F}{\d
  H^{\dagger}} + \left( \frac{\d\F}{\d Y_{\hat H}} \right)^T \left( \frac{\d\F}{\d \hat H}
\right)^T \bigg. \bigg] \nonumber\\
& & + s \e \frac{\partial\F}{\partial \e} +
  s\w^{\n} \frac{\partial\F}{\partial\w^{\n}} \nonumber\\
& &{} + \int{\rm d}^4x \, \bigg[ \bigg. sa
\frac{\d \F}{\d a} + sa^{\dagger}
 \frac{\d \F}{\d a^{\dagger}} + (s \Xi)^T
\left( \frac{\d \F}{\d \Xi} \right)^T \nonumber \\
& & \quad + s(f + f_0) \frac{\d \F}{\d f} + s(f^{\dagger} + f_0) \frac{\d \F}{\d f^{\dagger}} \bigg. \bigg]. 
\end{eqnarray}
The BRS transformations of the 4-spinors read
\begin{eqnarray}
s \Psi & = & \i ec\Psi -
\sqrt{2} y^{123} P_L \e \phi_{R}^{\dagger}(H^{\dagger} + v) 
+ \sqrt{2} y^{123} P_R \e \phi_{L} (H + v) \nonumber \\
& & + \sqrt{2} \i P_L \g^{\m} \e D_{\m}\phi_{L} -
 \sqrt{2} \i P_R \g^{\m} \e D_{\m}\phi_{R}^{\dagger} -
\i\w^{\n}\partial_{\n} \Psi, \\
(s \bar\Psi)^T & = & -\i ec{\bar\Psi}^T +
\sqrt{2} y^{123} P_R{\bar \e}^T \phi_{R} (H + v) -  
\sqrt{2} y^{123} P_L{\bar \e}^T \phi_{L}^{\dagger}(H^{\dagger} + v) \nonumber\\
& & + \sqrt{2} \i{\g^{\m}}^T P_L{\bar \e}^T D_{\m}\phi_{L}^{\dagger} -
\sqrt{2} \i{\g^{\m}}^T P_R{\bar \e}^T D_{\m}\phi_{R} -
\i\w^{\n}\partial_{\n}{\bar \Psi}^T, \\
s \tilde\g & = & \frac{\i}{4} [\g^{\sigma}, \g^{\rho}] \e F_{\rho
  \sigma} + (P_R - P_L) \e e Q_L (|\phi_L|^2 - |\phi_R|^2) \nonumber\\
& & + (P_R - P_L) \w^{\n}\partial_{\n} \tilde\g, \\
s \hat H & = & - \sqrt{2} P_L \e \left[y^{123}
  \phi_L^{\dagger} \phi_R^{\dagger} + \frac 12 y^{333} (H^{\dagger} + v)
  (H^{\dagger} + v) \right] \nonumber\\
& & + \sqrt{2} P_R \e \left[y^{123} \phi_L \phi_R + \frac 12 y^{333} (H + v)
  (H + v)\right] \nonumber\\
& & + \sqrt{2} \i \g^{\m} P_R \e \partial_{\m} H^{\dagger}
- \sqrt{2} \i \g^{\m} P_L \e \partial_{\m} H 
- \i\w^{\n}\partial_{\n} \hat H, \\ 
(s \Xi)^T & = & \sqrt 2 \hat f \e^T P_L - \sqrt 2{\hat f}^{\dagger} \e^T
P_R - \sqrt 2 \i \e^T{\g^{\m}}^T P_L \partial_{\m} a + \sqrt 2 \i \e^T
{\g^{\m}}^T P_R \partial_{\m} a^{\dagger} \nonumber\\
& & - \i \w^{\n} \partial_{\n} \Xi^T.
\end{eqnarray}

\section{One-loop results}
\label{app:results}

We list the one-loop diagrams of the vertex functions of identity 
(\ref{stiphys}) as well as the results given as one-loop integrals.
We choose $\xi = 1$ in the gauge fixing term and use the following abbreviations:
\begin{eqnarray}
d_{11} & = & \frac {1}{\sqrt 2 v} \left(2 m_e^2 + \frac {m_e m_H}{2} +
  \frac {M_1^2 - M_2^2}{2} \right), \\
d_{12} & = & \frac {1}{\sqrt 2 v} \left(2 m_e^2 - \frac {m_e m_H}{2} -
  \frac {M_1^2 - M_2^2}{2} \right), \\
d_{22} & = & \frac {1}{\sqrt 2 v} \left(- \frac 32 m_e m_H +
  \frac {M_1^2 - M_2^2}{2} \right) = - d_{21},\\
x_1 & = & \frac {1}{\sqrt 2 f_0} \left(M_1^2 - m_e^2 - \frac {m_e
    m_H}{2} \right),  \\
x_2 & = & \frac {1}{\sqrt 2 f_0} \left(M_2^2 - m_e^2 + \frac {m_e
    m_H}{2} \right), \\
x_{H_1} & = & \frac {1}{\sqrt 2 f_0} \left(M_{H_1}^2 - \frac 32 m_H^2 \right),\\
x_{H_2} & = & \frac {1}{\sqrt 2 f_0} \left(M_{H_2}^2 - \frac 12 m_H^2 \right),
\end{eqnarray}
and
\begin{equation}
\theta_{\rm DREG}=\left\{
\begin{array}{cl}
1 & \qquad \mbox{for DREG}, \\
0 & \qquad \mbox{for DRED}.
\end{array}
\right.
\end{equation}
For reasons of brevity we do not express $B_1,C_0,C_1$ integrals in terms of 
$A_0$ and $B_0$ integrals and omit the momentum arguments in
the one-loop functions (see Ref.\ \cite{Denner:kt} for definitions):
\begin{eqnarray}
C_{\{0,1\}}(m_0,m_1,m_2)&:=& C_{\{0,1\}}(p,0,m_0,m_1,m_2), \\
B_{\{0,1\}}(m_0,m_1)&:=& B_{\{0,1\}}(p,m_0,m_1).
\end{eqnarray}
Furthermore, we often combine the explicit results of two Green
functions into one equation where the upper sign of $\pm,\mp$ (first
index) corresponds to the first Green function and the lower one 
(second index) to the second.

\begin{figure}[t]
\setlength{\unitlength}{1cm}
\centerline{
\begin{picture}(8.0,2.5)
\put(-5,-21.6){\includegraphics{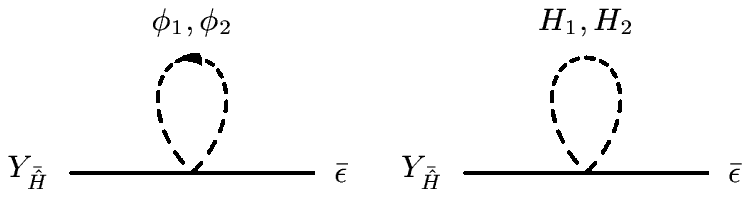}}
\end{picture}}
\caption{One-loop diagrams to $\Gamma_{Y_{\bar{\hat H}} \bar \e}$}
\end{figure}

\begin{figure}[t]
\setlength{\unitlength}{1cm}
\centerline{
\begin{picture}(5.0,3.6)
\put(-4,-20.3){\includegraphics{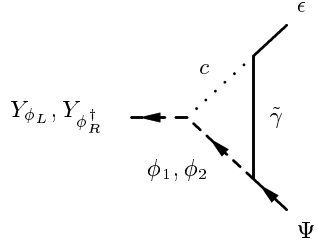}}
\end{picture}}
\caption{One-loop diagrams to $\Gamma_{\Psi \bar \e Y_{\phi_L}}$
and $\Gamma_{\Psi \bar \e Y_{\phi_R^{\dagger}}}$}
\end{figure}

\begin{figure}[t]
\setlength{\unitlength}{1cm}
\centerline{
\begin{picture}(16.0,6.0)
\put(-5.1,-18.0){\includegraphics{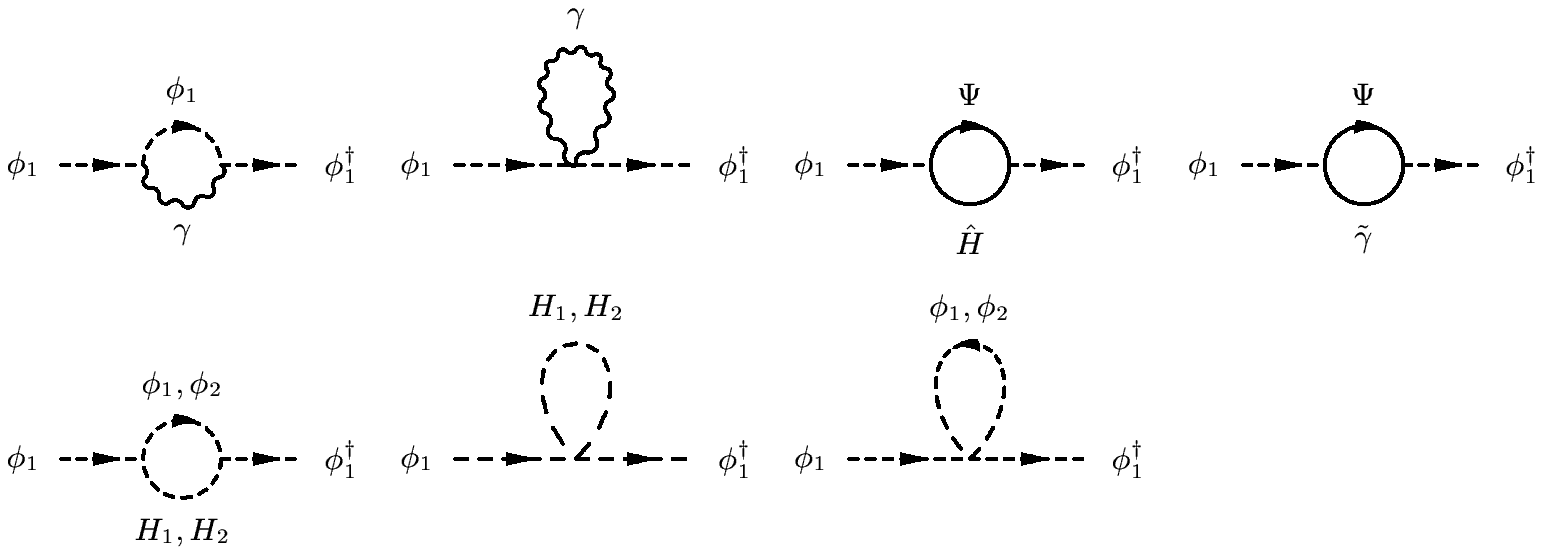}}
\end{picture}}
\caption{One-loop diagrams to $\Gamma_{\phi_1^{\dagger} \phi_1}$}
\end{figure}

\begin{figure}[t]
\setlength{\unitlength}{1cm}
\centerline{
\begin{picture}(16.0,6.0)
\put(-5.1,-18.0){\includegraphics{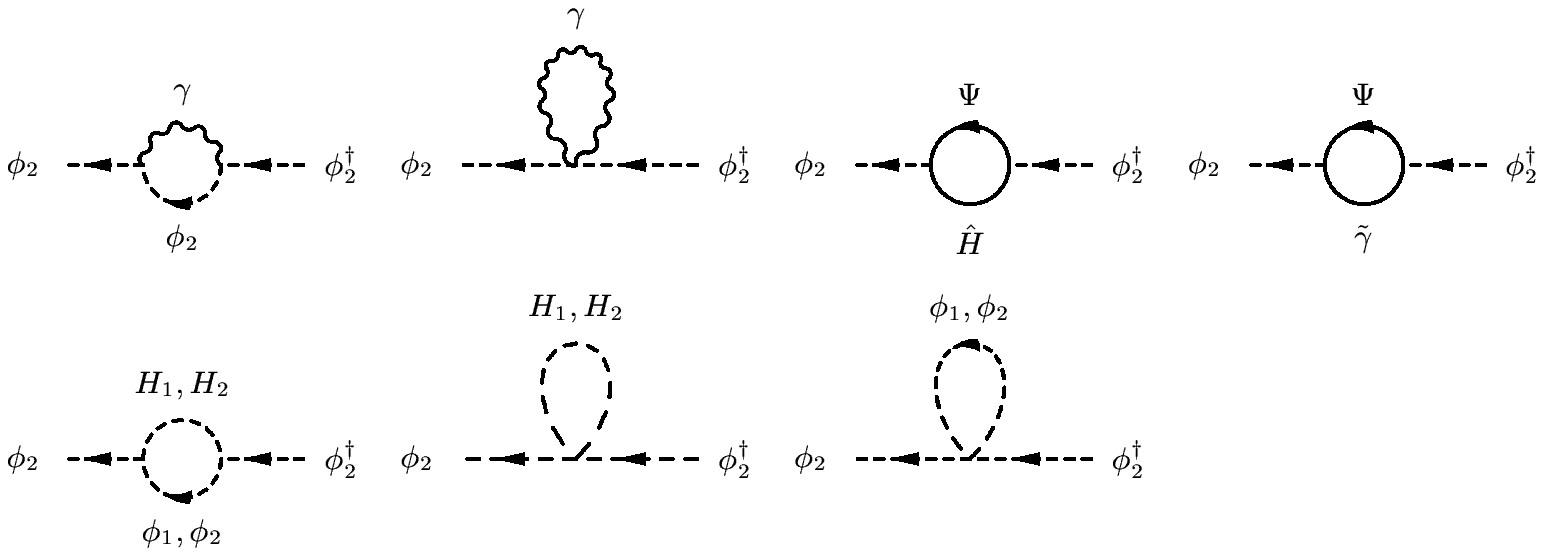}}
\end{picture}}
\caption{One-loop diagrams to $\Gamma_{\phi_2^{\dagger} \phi_2}$}
\end{figure}

\begin{figure}[t]
\setlength{\unitlength}{1cm}
\centerline{
\begin{picture}(14.0,8.0)
\put(-5.0,-16.0){\includegraphics{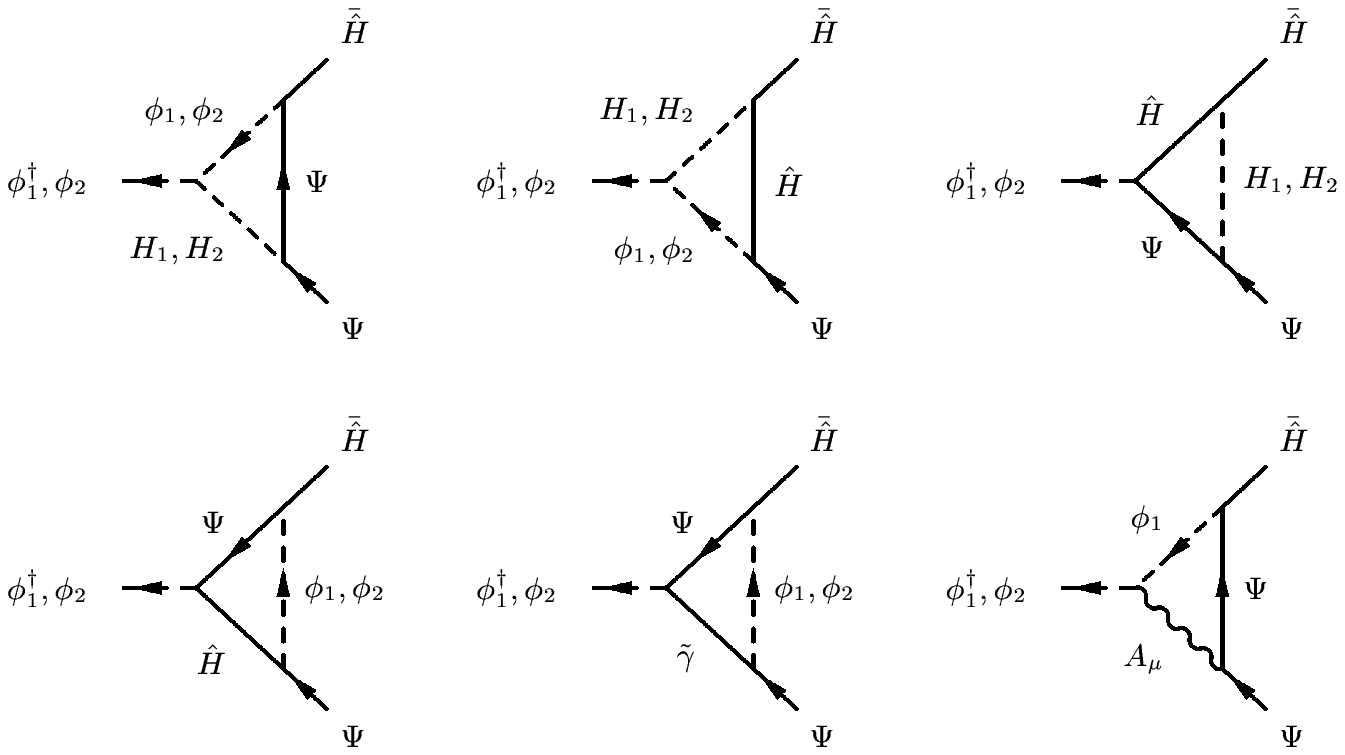}}
\end{picture}}
\caption{One-loop diagrams to 
$\Gamma_{\phi_1^{\dagger} \Psi \bar{\hat H}}$
and $\Gamma_{\phi_2 \Psi \bar{\hat H}}$}
\end{figure}

\begin{figure}[t]
\setlength{\unitlength}{1cm}
\centerline{
\begin{picture}(13.5,3.8)
\put(-5.7,-20.3){\includegraphics{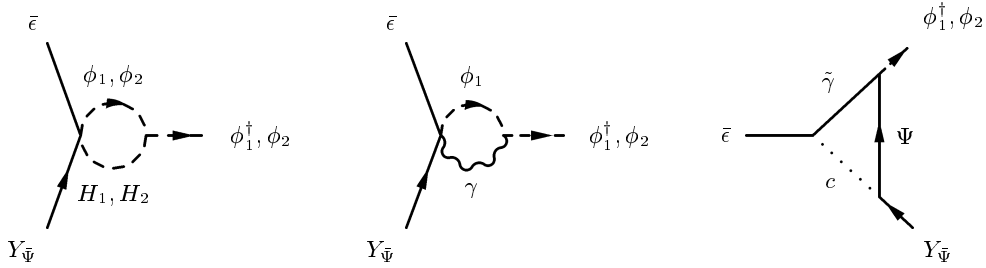}}
\end{picture}}
\caption{One-loop diagrams to 
$\Gamma_{\phi_1^{\dagger} Y_{\bar \Psi} \bar \e}$ and 
$\Gamma_{\phi_2 Y_{\bar \Psi} \bar \e}$}
\end{figure}

\begin{figure}[t]
\setlength{\unitlength}{1cm}
\centerline{
\begin{picture}(16.0,2.5)
\put(-5.2,-21.0){\includegraphics{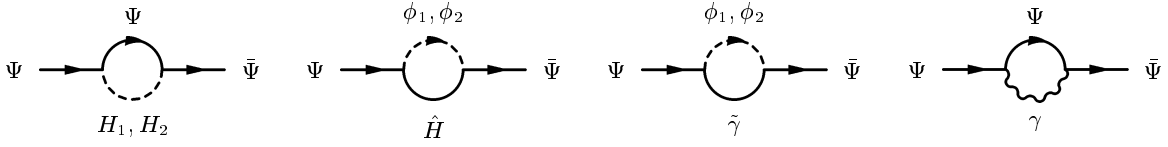}}
\end{picture}}
\caption{One-loop diagrams to $\Gamma_{\Psi \bar\Psi}^{(1)}$}
\end{figure}

\begin{figure}[t]
\setlength{\unitlength}{1cm}
\centerline{
\begin{picture}(14.0,15.0)
\put(-5.0,-9.0){\includegraphics{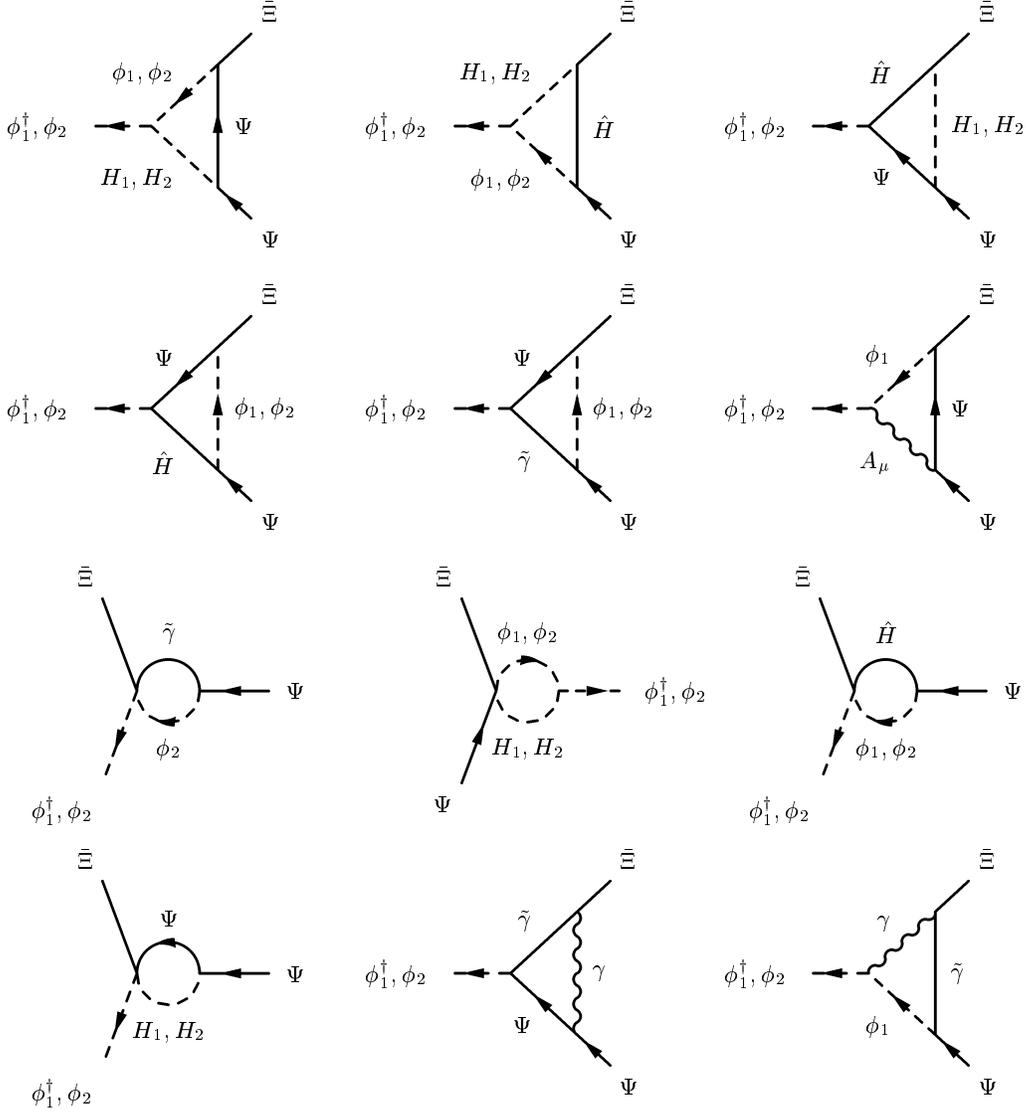}}
\end{picture}}
\caption{One-loop diagrams to 
$\Gamma_{\phi_1^{\dagger} \Psi \bar \Xi}$ and $\Gamma_{\phi_2 \Psi \bar\Xi}$}
\end{figure}

\begin{eqnarray}
\lefteqn{\Gamma_{Y_{\bar{\hat H}} \bar \e}^{(1)}(0, 0)
 = \frac{\a}{4 \pi} \frac{1}{\sqrt 2 e^2} \g_5 \big\{}
\nonumber \\ && {} 
y^{123} \left[ A_0 (M_2) - A_0 (M_1) \right] 
+y^{333} \left[ A_0 (M_{H_2}) - A_0 (M_{H_1})\right]\big\}, \\
\lefteqn{\Big\{\Gamma_{\Psi \bar \e Y_{\phi_L}}^{(1)} (p, 0, -p),
\Gamma_{\Psi \bar \e Y_{\phi_R^{\dagger}}}^{(1)} (p, 0, -p)\Big\}
= \frac{\a}{4 \pi} \frac{1}{\sqrt 2} \big\{}
\nonumber\\&& {}
- \g_5 \left[B_0 (m_{\tilde\g}, M_1) 
+ m_{\tilde\g} \, \pslash \, C_1 ( m_{\tilde\g}, M_1, 0)\right] 
\nonumber\\ && {}
\pm \left[ B_0 (m_{\tilde\g}, M_2) 
- m_{\tilde\g} \, \pslash \, C_1 ( m_{\tilde\g}, M_2, 0)\right]
\big\},\\
\lefteqn{\Big\{\Gamma_{\phi_1^{\dagger} \phi_1}^{(1)} (-p, p),
\Gamma_{\phi_2^{\dagger} \phi_2}^{(1)} (p, -p)\Big\}
  = \frac{\a}{4 \pi} \bigg\{}
\nonumber \\ && {} 
- \left[2 (M_{1,2}^2 + p^2) B_0 (0, M_{1,2}) - A_0(M_{1,2})\right] 
\nonumber \\ && {}
- \frac 12 \left( \frac{y^{123}}{e} \right)^2 \left[4 p^2 
  B_1 (m_{H,e}, m_{e,H}) + 4 A_0 (m_{e,H}) + 4(m_{H,e}^2 \pm m_e m_H) B_0 (m_H, m_e) \right]
\nonumber \\ && {}
- \left[4 p^2 B_1(m_{\tilde\g,e}, m_{e,\tilde\g}) + 4 A_0 (m_{e,\tilde\g}) + 
  4 \left(\mp m_e m_{\tilde\g} + m_{\tilde\g,e}^2\right) B_0 (m_{\tilde\g}, m_e) \right]
\nonumber \\ && {}
+ \frac{1}{e^2} \left[ d_{11,12}^2 B_0 (M_{H_1}, M_{1,2}) + d_{21}^2
  B_0 (M_{H_2}, M_{2,1}) \right] 
\nonumber \\ && {} 
+ \frac{1}{2 e^2} \left[\left(\left(y^{123}\right)^2 \pm \frac 12 y^{123}
  y^{333} \right) A_0 (M_{H_1}) + \left(\left(y^{123}\right)^2 \mp \frac
  12 y^{123} y^{333} \right) A_0 (M_{H_2}) \right] 
\nonumber \\ && {}
+ A_0 (M_{2,1}) + \left( \frac{y^{123}}{e} \right)^2 A_0 (M_{1,2})\bigg\}, \\
\lefteqn{\Big\{ \Gamma_{\phi_1^{\dagger} \Psi \bar{\hat H}}^{(1)} (-p,
  p, 0),\Gamma_{\phi_2 \Psi \bar{\hat H}}^{(1)} (-p, p, 0)\Big\}
  =  \frac{\a}{4 \pi} \{1,\g_5\} \bigg\{} 
\nonumber \\ && {}
  \pm \frac 12 \left( \frac{y^{123}}{e} \right)^2 \big[ 
- d_{11,12} \, \pslash \, C_1( m_e, M_{H_1}, M_{1,2}) 
+ d_{11,12} \, m_e \, C_0 ( m_e, M_{H_1}, M_{1,2}) 
\nonumber \\ && {}
\phantom{\mbox{} \pm \frac 12 \left( \frac{y^{123}}{e} \right)^2 \big[}
 \mp d_{21} \, \pslash \, C_1 ( m_e, M_{H_2}, M_{2,1}) \mp d_{21}
 \, m_e \, C_0 ( m_e, M_{H_2}, M_{2,1}) \big], 
\nonumber \\ && {}
\pm \frac 12 \frac{y^{123} y^{333}}{e^2} \big[ 
\mp d_{11,12} \, \pslash \, C_1( m_H, M_{1,2}, M_{H_1}) 
+ d_{11,12} \, m_H \, C_0 ( m_H, M_{1,2}, M_{H_1}) 
\nonumber \\ && {} 
\phantom{\mbox{}\pm \frac 12 \frac{y^{123} y^{333}}{e^2} \big[}
- d_{21} \, \pslash \, C_1 ( m_H, M_{2,1}, M_{H_2}) 
\mp d_{21} \, m_H \, C_0 ( m_H, M_{2,1}, M_{H_2}) \big] 
\nonumber \\ && {} 
\pm \frac{\left(y^{123}\right)^2 y^{333}}{2 \sqrt 2 \, e^2} \bigg[ 
\pm\left(p^2 + \pslash \left(m_e \pm m_H\right) \right) C_1 ( m_H, m_e, M_{H_1}) 
\nonumber \\ && {}
\phantom{\mbox{} \pm \, \frac{\left(y^{123}\right)^2 y^{333}}{2 \sqrt 2 \, e^2} \bigg[}
\pm \frac{1}{M_{H_1}^2 - m_H^2} \left( M_{H_1}^2 \, 
B_0 (M_{H_1}, m_e) -  m_H^2 \, B_0 (m_H, m_e) \right) 
\nonumber \\ && {}
\phantom{\mbox{} \pm \, \frac{\left(y^{123}\right)^2 y^{333}}{2 \sqrt 2 \, e^2} \bigg[}
+ \left( \pslash m_H + m_e m_H\right) \, C_0( m_H, m_e, M_{H_1})
\nonumber \\ && {}
\phantom{\mbox{} \pm \, \frac{\left(y^{123}\right)^2 y^{333}}{2 \sqrt 2 \, e^2} \bigg[}
\mp \left(p^2 - \pslash \left(m_e \pm m_H\right) \right) C_1 ( m_H, m_e,M_{H_2})
\nonumber \\ && {}
\phantom{\mbox{} \pm \, \frac{\left(y^{123}\right)^2 y^{333}}{2 \sqrt 2 \, e^2} \bigg[}
\mp \frac{1}{M_{H_2}^2 - m_H^2} \left( M_{H_2}^2 \, 
B_0 (M_{H_2}, m_e) -  m_H^2 \, B_0 ( m_H, m_e) \right) 
\nonumber \\ && {}
\phantom{\mbox{} \pm \, \frac{\left(y^{123}\right)^2 y^{333}}{2 \sqrt 2 \, e^2} \bigg[}
- \left(m_e m_H - \pslash m_H\right) \, C_0 ( m_H, m_e, M_{H_2})\bigg] 
\nonumber \\ && {} 
\pm \frac{\left(y^{123}\right)^3}{2 \sqrt 2 \, e^2} \bigg[
\pm\left(p^2 + \pslash \left(m_H \pm m_e\right) \right) C_1 ( m_e, m_H,M_1) 
\nonumber \\ && {}
\phantom{\mbox{}\pm \, \frac{\left(y^{123}\right)^3}{2 \sqrt 2 \, e^2} \bigg[}
\pm \frac{1}{M_1^2 - m_e^2} \left( M_1^2 \, B_0 (M_1, m_H) 
-  m_e^2 \, B_0 ( m_e, m_H) \right) 
\nonumber \\ && {}
\phantom{\mbox{}\pm \, \frac{\left(y^{123}\right)^3}{2 \sqrt 2 \, e^2} \bigg[}
+ \left( \pslash m_e + m_e m_H\right) \, C_0( m_e, m_H, M_1)
\nonumber \\ && {}
\phantom{\mbox{}\pm \, \frac{\left(y^{123}\right)^3}{2 \sqrt 2 \, e^2} \bigg[}
\mp \left(p^2 - \pslash \left(m_H \pm m_e\right) \right) C_1 ( m_e, m_H, M_2)
\nonumber \\ && {}
\phantom{\mbox{}\pm \, \frac{\left(y^{123}\right)^3}{2 \sqrt 2 \, e^2} \bigg[}
\mp \frac{1}{M_2^2 - m_e^2} \left( M_2^2 \, B_0 (M_2, m_H) 
-  m_e^2 \, B_0 (m_e, m_H) \right) 
\nonumber \\ && {}
\phantom{\mbox{}\pm \, \frac{\left(y^{123}\right)^3}{2 \sqrt 2 \, e^2} \bigg[}
- \left(m_e m_H - \pslash m_e\right) \, C_0 ( m_e, m_H, M_2)\bigg] 
\nonumber \\ && {} 
+ \frac{y^{123}}{\sqrt 2} \bigg[ 
\mp \left( p^2 - \pslash \left(m_{\tilde\g} \mp m_e\right) \right) 
C_1 (m_e, m_{\tilde\g}, M_1) 
\nonumber \\ && {}
\phantom{\mbox{}+ \frac{y^{123}}{\sqrt 2} \Big[}
\mp \frac{1}{M_1^2 - m_e^2} \left( M_1^2 \, B_0 (M_1, m_{\tilde\g}) 
-  m_e^2 \, B_0 (m_e, m_{\tilde\g}) \right) 
\nonumber \\ && {}
\phantom{\mbox{}+ \frac{y^{123}}{\sqrt 2} \Big[}
+ \left(- \pslash m_e + m_e m_{\tilde\g}\right) \, C_0( m_e, m_{\tilde\g}, M_1)
\nonumber \\ && {}
\phantom{\mbox{}+ \frac{y^{123}}{\sqrt 2} \Big[}
\mp \left(p^2 + \pslash \left(m_{\tilde\g} \mp m_e\right) \right) 
C_1 (m_e, m_{\tilde\g}, M_2) 
\nonumber \\ && {}
\phantom{\mbox{}+ \frac{y^{123}}{\sqrt 2} \Big[}
\mp \frac{1}{M_2^2 - m_e^2} \left( M_2^2 \, B_0 (M_2, m_{\tilde\g}) 
-  m_e^2 \, B_0 (m_e, m_{\tilde\g}) \right) 
\nonumber \\ && {}
\phantom{\mbox{}+ \frac{y^{123}}{\sqrt 2} \Big[}
+ \left(m_e m_{\tilde\g} + \pslash m_e\right) \, C_0 ( m_e, m_{\tilde\g},M_2) \bigg]
\nonumber \\ && {}
\mp \, \frac{y^{123}}{\sqrt 2} \bigg[ 
\frac 12 \left( B_0 (M_{1,2}, 0) + B_0 (m_e, 0) 
+ \left(m_e^2 + M_{1,2}^2\right) \, C_0 ( m_e, 0, M_{1,2})\right) 
\nonumber \\ && {}
\phantom{\mbox{}\mp \, \frac{y^{123}}{\sqrt 2} \bigg[}
+ \pslash m_e \left(C_0( m_e, 0, M_{1,2}) - C_1( m_e, 0, M_{1,2})\right) 
\nonumber \\ && {}  
\phantom{\mbox{}\mp \, \frac{y^{123}}{\sqrt 2} \bigg[}
- \, p^2 C_1 ( m_e, 0, M_{1,2}) \bigg] \bigg\}, \\
\lefteqn{\Big\{ \Gamma_{\phi_1^{\dagger} Y_{\bar \Psi} \bar \e}^{(1)} 
(-p, p, 0),\Gamma_{\phi_2 Y_{\bar \Psi} \bar \e}^{(1)} (-p, p, 0)\Big\}
  =  \frac{\a}{4 \pi} \{ \g_5,1\} \bigg\{} 
\nonumber \\ && {}
\pm\frac{y^{123}}{\sqrt 2 \, e^2} \left[ d_{21} \, 
B_0 (M_{H_2}, M_{2,1}) \mp d_{11,12} \, B_0 (M_{H_1}, M_{1,2})\right] 
\nonumber \\ && {}
- \left[ 2 \pslash B_0(0, M_{1,2}) + \pslash B_1 (0, M_{1,2} )\right] 
\nonumber \\ && {}
\mp \big[ \mp \pslash \, B_1(m_{\tilde\g}, m_e) \mp 
\left(\pslash + m_e \mp m_{\tilde\g}\right) \, B_0 (m_{\tilde\g}, m_e) 
\nonumber \\ && {}
\phantom{\mbox{}\mp \big[}
+ \left(m_{\tilde\g} p^2 + m_{\tilde\g} m_e \pslash\right) \, 
C_1 (m_{\tilde\g}, m_e, 0)\big] \bigg\},\\
\lefteqn{\Gamma_{\Psi\bar\Psi}^{(1)}(p,-p)=\frac{\a}{4 \pi} \bigg\{} 
\nonumber \\ && {}
\frac 12 \left(\frac{y^{123}}{e}\right)^2 \big[ 
\pslash \left(B_0 (M_{H_1}, m_e) + B_1 (M_{H_1}, m_e)\right) 
+ m_e B_0 (M_{H_1}, m_e) 
\nonumber \\ && {}
\phantom{\mbox{}\frac 12 \left(\frac{y^{123}}{e}\right)^2 \big[}
+ \pslash \left(B_0 (M_{H_2}, m_e) + B_1 (M_{H_2}, m_e)
  \right) - m_e B_0 (M_{H_2}, m_e) \big] 
\nonumber \\ && {}
+ \frac 12 \left(\frac{y^{123}}{e}\right)^2 \big[ 
\pslash \left(B_0 (M_1, m_H) + B_1 (M_1, m_H)\right) 
+ m_H B_0 (M_1, m_H) 
\nonumber \\ && {}
\phantom{\mbox{}+ \frac 12 \left(\frac{y^{123}}{e}\right)^2 \big[}
+ \pslash \left(B_0 (M_2, m_H) + B_1 (M_2, m_H)\right) 
- m_H B_0 (M_2, m_H) 
\big] 
\nonumber \\ && {}
+ \big[ \pslash \left(B_0 (M_1, m_{\tilde\g}) + B_1 (M_1, m_{\tilde\g})\right) 
- m_{\tilde\g} B_0 (M_1, m_{\tilde\g}) 
\nonumber \\ && {}
\phantom{\mbox{}+ \big[}
+ \pslash \left(B_0 (M_2, m_{\tilde\g}) + B_1 (M_2, m_{\tilde\g})\right)
+ m_{\tilde\g} B_0 (M_2, m_{\tilde\g}) \big] 
\nonumber \\ && {}
- \left[ \left(4 m_e - \pslash - \pslash \frac{m_e^2}{p^2} \right)
  B_0 (0, m_e) + \frac{\pslash}{p^2} A_0(m_e) + \theta_{\rm DREG}
  \left(\pslash - 2 m_e\right) \right] \bigg\}, \\
\lefteqn{\Big\{\Gamma_{\phi_1^{\dagger} \Psi \bar \Xi}^{(1)} (-p, p, 0),
\Gamma_{\phi_2 \Psi \bar \Xi}^{(1)} (-p, p, 0)\Big\}
  =  \frac{\a}{4 \pi} \{1,\g_5\} \bigg \{}
\nonumber \\ && {}
- \frac{y^{123}}{\sqrt 2 \, e^2} \big[ 
\pm x_1 \, d_{11,21} \, \left(\mp \pslash \, 
  C_1( m_e, M_{H_{1,2}}, M_1) + m_e \, C_0 ( m_e, M_{H_{1,2}}, M_1)\right) 
\nonumber \\ && {}
\phantom{\mbox{}- \frac{y^{123}}{\sqrt 2 \, e^2} \big[}
+ x_2 \, d_{21,12} \, \left(\pm\pslash \, C_1 ( m_e, M_{H_{2,1}}, M_2) +
  m_e \, C_0 ( m_e, M_{H_{2,1}}, M_2) \right) \big] 
\nonumber \\ && {} 
- \frac{y^{123}}{\sqrt 2 \, e^2} \big[ 
- x_{H_1} \, d_{11,12} \, \left(\pslash \, C_1( m_H, M_{1,2}, M_{H_1})
\mp m_H \, C_0 ( m_H, M_{1,2}, M_{H_1}) \right) 
\nonumber \\ && {} 
\phantom{\mbox{}- \frac{y^{123}}{\sqrt 2 \, e^2} \big[}
+ x_{H_2} \, d_{21} \, \left(\pm\pslash \, C_1 ( m_H, M_{2,1}, M_{H_2})
+ m_H \, C_0 ( m_H, M_{2,1}, M_{H_2}) \right) \big] 
\nonumber \\ && {}
\mp \frac 12 \left(\frac{y^{123}}{e} \right)^2 \bigg[ x_{H_1} \bigg[ 
\left(\pm p^2 + \pslash (m_H \pm m_e) \right) 
  C_1 ( m_H, m_e, M_{H_1}) 
\nonumber \\ && {} 
\phantom{\mbox{}\mp \frac 12 \left(\frac{y^{123}}{e} \right)^2 \bigg[ x_{H_1} \bigg[}
\pm \frac{1}{M_{H_1}^2 - m_H^2} \left( M_{H_1}^2 \, B_0 (M_{H_1}, m_e) 
- m_H^2 \, B_0 ( m_H, m_e) \right) 
\nonumber \\ && {} 
\phantom{\mbox{}\mp \frac 12 \left(\frac{y^{123}}{e} \right)^2 \bigg[ x_{H_1} \bigg[}
+ \left( \pslash m_H + m_e m_H\right) \, C_0( m_H, m_e, M_{H_1}) \bigg] 
\nonumber \\ && {}
\phantom{\mbox{}\mp \frac 12 \left(\frac{y^{123}}{e} \right)^2 \bigg[} 
+ x_{H_2} \bigg[
\left(\pm p^2 - \pslash \left(m_H \pm m_e\right) \right) C_1 ( m_H, m_e, M_{H_2}) 
\nonumber \\ && {}
\phantom{\mbox{}\mp \frac 12 \left(\frac{y^{123}}{e} \right)^2 \bigg[+ x_{H_2} \bigg[} 
\pm \, \frac{1}{M_{H_2}^2 - m_H^2} \left( M_{H_2}^2 \, B_0 (M_{H_2},m_e) 
-  m_H^2 \, B_0 ( m_H, m_e) \right) 
\nonumber \\ && {}
\phantom{\mbox{}\mp \frac 12 \left(\frac{y^{123}}{e} \right)^2 \bigg[+ x_{H_2} \bigg[} 
- \left(\pslash m_H - m_e m_H\right) \, C_0 ( m_H, m_e, M_{H_2})\big] \bigg] 
\nonumber \\ && {} 
\mp \frac 12 \left( \frac{y^{123}}{e} \right)^2 \bigg[ 
x_1 \bigg[ \pm\left(p^2 + \pslash \left(m_H \pm m_e\right) \right) C_1 (m_e, m_H,M_1) 
\nonumber \\ && {}
\phantom{\mbox{}\mp \frac 12 \left( \frac{y^{123}}{e} \right)^2 \bigg[ x_1 \bigg[} 
\pm \frac{1}{M_1^2 - m_e^2} \left( M_1^2 \, B_0 (M_1, m_H) 
-  m_e^2 \, B_0 ( m_e, m_H) \right) 
\nonumber \\ && {}
\phantom{\mbox{}\mp \frac 12 \left( \frac{y^{123}}{e} \right)^2 \bigg[ x_1 \bigg[} 
+ \left( \pslash m_e + m_e m_H\right) \, C_0( m_e, m_H, M_1)\bigg] 
\nonumber \\ && {}
\phantom{\mbox{}\mp \frac 12 \left( \frac{y^{123}}{e} \right)^2 \bigg[} 
+ x_2 \bigg[ \pm\left(p^2 - \pslash \left(m_H \pm m_e\right) \right) C_1 ( m_e, m_H,M_2)
\nonumber \\ && {}
\phantom{\mbox{}\mp \frac 12 \left( \frac{y^{123}}{e} \right)^2 \bigg[+ x_2 \bigg[} 
\pm \frac{1}{M_2^2 - m_e^2} \left( M_2^2 \, B_0 (M_2, m_H) 
-  m_e^2 \, B_0 ( m_e, m_H) \right) 
\nonumber \\ && {}
\phantom{\mbox{}\mp \frac 12 \left( \frac{y^{123}}{e} \right)^2 \bigg[+ x_2 \bigg[} 
+ \left(m_e m_H - \pslash m_e\right) \, C_0 ( m_e, m_H, M_2) \bigg] \bigg] 
\nonumber \\ && {} 
- \bigg[ x_1 \bigg[ \mp \left(p^2 - \pslash \left(m_{\tilde\g} \mp m_e\right) \right)
C_1 ( m_e, m_{\tilde\g}, M_1) 
\nonumber \\ && {}
\phantom{\mbox{}- \bigg[x_1 \bigg[}
\mp \frac{1}{M_1^2 - m_e^2} \left( M_1^2 \, B_0 (
 M_1, m_{\tilde\g}) -  m_e^2 \, B_0 ( m_e, m_{\tilde\g}) \right) 
\nonumber \\ && {}
\phantom{\mbox{}- \bigg[x_1 \bigg[}
+ \left(- \pslash m_e + m_e m_{\tilde\g}\right) \, C_0( m_e,m_{\tilde\g}, M_1) \bigg] 
\nonumber \\ && {}
\phantom{\mbox{}- \bigg[}
- x_2 \bigg[ \mp \, \left(p^2 + \pslash \left(m_{\tilde\g} \mp m_e\right)\right) 
C_1 ( m_e, m_{\tilde\g}, M_2) 
\nonumber \\ && {}
\phantom{\mbox{}- \bigg[- x_2 \bigg[}
\mp \frac{1}{M_2^2 - m_e^2} \left( M_2^2 \, B_0 (M_2, m_{\tilde\g}) 
-  m_e^2 \, B_0 ( m_e, m_{\tilde\g}) \right) 
\nonumber \\ && {}
\phantom{\mbox{}- \bigg[- x_2 \bigg[}
+ \, \left(m_e m_{\tilde\g} + \pslash m_e\right) \, C_0 ( m_e, m_{\tilde\g},M_2) 
\big] \bigg] 
\nonumber \\ && {}
+ x_{1,2} \bigg[ \frac 12 \left( B_0 (M_{1,2}, 0) + B_0 (m_e, 0) + \left(m_e^2 + M_{1,2}^2\right)
\, C_0 ( m_e, 0, M_{1,2}) \right) 
\nonumber \\ && {}
\phantom{\mbox{}+ x_{1,2} \bigg[}
+ \pslash m_e \left(C_0( m_e, 0, M_{1,2}) - C_1( m_e, 0, M_{1,2})\right) 
\nonumber \\ && {}  
\phantom{\mbox{}+ x_{1,2} \bigg[}
- p^2 C_1 ( m_e, 0, M_{1,2}) \bigg] 
\nonumber \\ && {}
+ \frac{m_{\tilde\g}}{\sqrt 2 \, f_0} \left[ 
\mp \pslash B_1( m_{\tilde\g}, M_{2,1}) + m_{\tilde\g} 
B_0 ( m_{\tilde\g},M_{2,1}) \right] 
\nonumber \\ && {}
+ \frac{3{\tilde A}_{123}}{e^2} \left[ d_{21} \,
B_0 ( M_{2,1}, M_{H_2})+ \{- d_{11},d_{12}\} \, B_0 ( M_{1,2}, M_{H_1})\right] 
\nonumber \\ && {} 
\mp \frac{3{\tilde A}_{123} \, y^{123}}{\sqrt 2 \, e^2} \big[ 
- \pslash \, B_1 ( m_H, M_{1,2}) \pm m_H \, B_0 (m_H, M_{1,2}) 
\nonumber \\ && {}
\phantom{\mbox{}\mp \frac{3{\tilde A}_{123} \, y^{123}}{\sqrt 2 \, e^2} \big[}
\mp \pslash B_1 ( m_H, M_{2,1}) \mp m_H \, B_0 (m_H, M_{2,1}) \big] 
\nonumber \\ && {}
+ \frac{3{\tilde A}_{123} \, y^{123}}{\sqrt 2 \, e^2} \big[ 
\mp \left(\pslash + m_e\right) \, B_0 ( M_{H_1}, m_e) \mp \pslash \,
B_1 ( M_{H_1}, m_e) 
\nonumber \\ && {}
\phantom{\mbox{}+ \frac{3{\tilde A}_{123} \, y^{123}}{\sqrt 2 \, e^2} \big[}
\mp \left(\pslash - m_e\right) B_0 ( M_{H_2}, m_e) \mp \pslash \, 
B_1 ( M_{H_2}, m_e) \big] 
\nonumber \\ && {}
+ \frac{m_{\tilde\g}}{\sqrt 2 \, f_0} \big[
\mp \frac{\pslash}{m^2_{\tilde\g}} \left[\left(p^2 - m_e^2 + m_{\tilde\g}^2\right) \, 
B_1 ( m_{\tilde\g}, m_e) - (p^2 - m_e^2) \, B_1( 0, m_e)\right] 
\nonumber \\ && {}
\phantom{\mbox{}+ \frac{m_{\tilde\g}}{\sqrt 2 \, f_0} \big[}
+ 3 m_{\tilde\g} \, \left(p^2 - \pslash m_e\right) C_1 (m_{\tilde\g} m_e, 0) 
\nonumber \\ && {}
\phantom{\mbox{}+ \frac{m_{\tilde\g}}{\sqrt 2 \, f_0} \big[}
+ \left(\pm\pslash + 3 \left(m_{\tilde\g} \mp m_e\right)\right) B_0 (m_{\tilde\g}, m_e) 
\nonumber \\ && {}
\phantom{\mbox{}+ \frac{m_{\tilde\g}}{\sqrt 2 \, f_0} \big[}
\pm 3 \pslash \, B_1 ( m_{\tilde\g}, m_e)
+ \theta_{\rm DREG} \left(-2 (m_{\tilde\g} \mp m_e\right) \mp \pslash) \big] 
\nonumber \\ && {}
\pm \frac{m_{\tilde\g}}{\sqrt 2 \, f_0} \bigg[ 
- \frac{\pslash}{m^2_{\tilde\g}} \left[\left(p^2 - M_{1,2}^2 + m_{\tilde\g}^2\right) \, 
B_1 ( m_{\tilde\g}, M_{1,2}) - (p^2 - M_{1,2}^2) \, 
B_1( 0, M_{1,2})\right] 
\nonumber \\ && {}
\phantom{\mbox{}\pm \frac{m_{\tilde\g}}{\sqrt 2 \, f_0} \big[} 
- 2 \pslash \, B_0 ( m_{\tilde\g}, M_{1,2}) \bigg] \bigg\}.
\end{eqnarray}

\end{appendix}

\section*{Acknowledgement}

We like to thank H.\ Rzehak for further informations about the results 
of Ref.\ \cite{Heidi}.

\end{document}